\documentclass[english, 12pt]{article}
\usepackage{enumerate}
\usepackage{amsopn,amsfonts,amsmath,amssymb,mathrsfs}
\usepackage[pdftex]{graphicx}
\usepackage{verbatim}
\usepackage[width=\textwidth]{caption}
\usepackage{booktabs}
\usepackage[english]{babel}
\usepackage{subfigure}
\usepackage[]{float}
\usepackage[]{placeins}
\usepackage{flafter}
\usepackage[]{longtable}
\usepackage{multirow}
\usepackage{verbatim}
\usepackage{dsfont}
\usepackage{array}	
\usepackage{natbib}
\usepackage{xr}
\usepackage{color}
\usepackage[usenames,dvipsnames,svgnames,table]{xcolor}
\usepackage[colorlinks=true,	
            linkcolor=red,
            urlcolor=blue,
            citecolor=gray]{hyperref}
\usepackage{layout}
\usepackage[top=1.25in, bottom=1.25in, left=1.25in, right=1.25in]{geometry}

\newcommand{\eps}{\varepsilon}
\def\E{{\mathbb E}}
\def\indic{{\rm {\large 1}\hspace{-2.3pt}{\large l}}}

\def\R{{\mathbb R}}

\def\N{{\mathbb N}}

\newcommand\indep{\protect\mathpalette{\protect\independenT}{\perp}}
\def\independenT#1#2{\mathrel{\rlap{$#1#2$}\mkern2mu{#1#2}}}

\def\F{{\mathcal F}}

\def\argmin{{\rm argmin}}

\newcommand{\df}[2]{\frac{#1 }{#2}}

\newcommand{\norm}[1]{\|#1 \|}

\newcommand{\abs}[1]{ \left|  #1\right| }
\newcommand{\mt}[1]{ \boldsymbol{ #1 } }

\newcommand{\Cov}{\text{Cov}}
\newcommand{\V}{\mathbb{V}}

\newcommand{\Supp}{\text{Supp}}

\newcommand{\RE}{\E\left[Y\middle|\mathcal{I}\right]}

\newcommand{\Yh}{\widehat{Y}}
\newcommand{\psh}{\widehat{\psi}}

\newtheorem{theorem}{Theorem}
\newtheorem{proposition}{Proposition}
\newtheorem{lemma}{Lemma}

\newtheorem{assumption}{Assumption}

\newcolumntype{C}[1]{>{\centering\let\newline\\\arraybackslash\hspace{0pt}}m{#1}}

\date{}

\begin{document}

\title{\vspace{-1.5cm}
Rationalizing Rational Expectations: Characterizations and Tests\thanks{This paper is based on portions of our previous working paper \cite{DGM}. We thank the co-editor Andres Santos, three anonymous referees, Peter Arcidiacono, Levon Barseghyan, Federico Bugni, Pierre Cahuc, Zhuoli Chen, Tim Christensen, Valentina Corradi, Christian Gourieroux, Nathael Gozlan, Gregory Jolivet, Max Kasy, Jia Li, Matt Masten, Magne Mogstad, Andrew Patton, Aureo de Paula, Mirko Wiederholt, Basit Zafar, Yichong Zhang and participants of various seminars and conferences for useful comments and suggestions.}}

\author{Xavier D'Haultfoeuille\thanks{CREST-ENSAE, xavier.dhaultfoeuille@ensae.fr. Xavier D'Haultfoeuille thanks the hospitality of PSE where part of this research was  conducted.} \and Christophe Gaillac\thanks{CREST-ENSAE and TSE, christophe.gaillac@tse-fr.eu.} \and Arnaud Maurel\thanks{Duke University, NBER and IZA, arnaud.maurel@duke.edu.}}
\maketitle
~\vspace{-1cm}
\begin{abstract}
In this paper, we build a new test of rational expectations based on the marginal distributions of realizations and subjective beliefs. This test is widely applicable, including in the common situation where realizations and beliefs are observed in two different datasets that cannot be matched. We show that whether one can rationalize rational expectations is equivalent to the distribution of realizations being a mean-preserving spread of the distribution of beliefs. The null hypothesis can then be rewritten as a system of many moment inequality and equality constraints, for which tests have been recently developed in the literature. The test is robust to measurement errors under some restrictions and can be extended to account for aggregate shocks. Finally, we apply our methodology to test for rational expectations about future earnings. While individuals tend to be right on average about their future earnings, our test strongly rejects rational expectations.


\end{abstract}

\newpage
\section{Introduction}\label{sec1}

How individuals form their beliefs about uncertain future outcomes is critical to understanding decision making. Despite longstanding critiques \citep*[see, among many others,][]{pesaran1987limits,Manski2004}, rational expectations remain by far the most popular framework to describe belief formation \citep{Muth61}. This theory states that agents have expectations that do not systematically differ from the realized outcomes, and efficiently process all private information to form these expectations. Rational expectations (RE) are a key building block in many macro- and micro-economic models, and in particular in most of the dynamic microeconomic models that have been estimated over the last two decades \citep[see, e.g.,][for recent surveys]{AM10,blundell17}.

\medskip
In this paper, we build a new test of RE. Our test only requires having access to the marginal distributions of subjective beliefs and realizations, and, as such, can be applied quite broadly. In particular, this test can be used in a data combination context, where individual realizations and subjective beliefs are observed in two different datasets that cannot be matched. Such situations are common in practice \citep*[see, e.g.,][]{Delavande2008, AHK2012, AHMR2014, Stinebrickner14b,gennaioli2016expectations,kuchler2015personal,BR18,BBRR20}. Besides, even in surveys for which an explicit aim is to measure subjective expectations, such as the Michigan Survey of Consumers or the Survey of Consumer Expectations of the New York Fed, expectations and realizations can typically only be matched for a subset of the respondents. And of course, regardless of attrition, whenever one seeks to measure long or medium-term outcomes, matching beliefs with realizations does require waiting for a long period of time before the data can be made available to researchers.\footnote{Situations where realizations can be perfectly predicted beforehand, such as in school choice settings where assignments are a known function of observed inputs, are notable exceptions.}

\medskip
The tests of RE implemented so far in this context  only use specific implications of the RE hypothesis. In contrast, we develop a test that exploits all possible implications of RE. Using the key insight that we can rationalize RE if and only if the distribution of realizations is a mean-preserving spread of the distribution of beliefs, we show that rationalizing RE is equivalent to satisfying one moment equality and (infinitely) many moment inequalities.\footnote{Interestingly, the equivalence on which we rely, which is based on Strassen's theorem \citep{strassen1965}, is also used in the microeconomic risk theory literature, see in particular \cite{rothschild1970increasing}.} As a consequence, if these moment conditions hold, RE cannot be rejected,  given the data at our disposal. By exhausting all relevant implications of RE, our test is able to detect much more violations of rational expectations than existing tests.

\medskip
To develop a statistical test of RE rationalization, we build on the recent literature on inference based on moment inequalities, and more specifically, on \cite{andrews2016inference}. By applying their results to our context, we show that our test controls size asymptotically and is consistent over fixed alternatives. We also provide conditions under which the test is not conservative.

\medskip
We then consider several extensions to our baseline test. First, we show that by using a set of covariates that are common to both datasets, we can increase our ability to detect violations of RE. Another important issue is that of unanticipated aggregate shocks. Even if individuals have rational expectations, the mean of observed outcomes may differ from the mean of individual beliefs simply because of aggregate shocks. We show that our test can be easily adapted to account for such shocks. 

\medskip
Finally, we prove that our test is robust to measurement errors in the following sense. If individuals have rational expectations but both beliefs and outcomes are measured with (classical) errors, then our test does not reject RE provided that the amount of measurement errors on beliefs does not exceed the amount of intervening transitory shocks plus the measurement errors on the realized outcomes. In that specific sense, imperfect data quality does not jeopardize the validity of our test. In particular, this allows for elicited beliefs to be noisier than realized outcomes. This provides a rationale for our test even in cases where realizations and beliefs are observed in the same dataset, since a direct test based on a regression of the outcome on the beliefs (see, e.g., \citeauthor{Lovell86}, \citeyear{Lovell86}) is, at least at the population level, not robust to any amount of measurement errors on the subjective beliefs.

\medskip
We apply our framework to test for rational expectations about future earnings. To do so, we combine elicited beliefs about future earnings with realized earnings, using data from the Labor Market module of the Survey of Consumer Expectations (SCE, New York Fed), and test whether household heads form rational expectations on their annual labor earnings. While a naive test of equality of means between earnings beliefs and realizations shows that earnings expectations are realistic in the sense of not being significantly biased, thus not rejecting the rational expectations hypothesis, our test does reject rational expectations at the 1\% level. Taken together, our findings illustrate the practical importance of incorporating the additional restrictions of rational expectations that are embedded in our test. The results of our test also indicate that the RE hypothesis is more credible for certain subpopulations than others. For instance, we reject RE for individuals without a college degree, who exhibit substantial deviations from RE. On the other hand, we fail to reject the hypothesis that college-educated workers have rational expectations on their future earnings.

\medskip
By developing a test of rational expectations in a setting where realizations and subjective beliefs are observed in two different datasets, we bring together the literature on data combination (see, e.g., \citeauthor{CM02}, \citeyear{CM02}, \citeauthor{MP06}, \citeyear{MP06}, \citeauthor*{fan2014identifying}, \citeyear{fan2014identifying}, \citeauthor*{BLL16}, \citeyear{BLL16}, and \citeauthor{RM2007}, \citeyear{RM2007} for a survey), and the literature on testing for rational expectations in a micro environment \citep[see, e.g.,][for seminal contributions]{Lovell86,gourieroux1986direct,ivaldi1992survey}.

\medskip
On the empirical side, we contribute to a rapidly growing literature on the use of subjective expectations data in economics \citep*[see, e.g.,][]{Manski2004,Delavande2008,KW2008,vanderKlaauw2012,AHMR2014,PST2014, Stinebrickner14,WZ2015}. In this paper, we show how to incorporate all of the relevant information from subjective beliefs combined with realized data to test for rational expectations. 

\medskip
The remainder of the paper is organized as follows. In Section \ref{sec2}, we present the general set-up and the main theoretical equivalences underlying our RE test. In Section \ref{sub:stat_tests}, we introduce the corresponding statistical tests and study their asymptotic properties. Section \ref{ss32} illustrates the finite sample properties of our tests through Monte Carlo simulations. Section \ref{sec4} applies our framework to expectations about future earnings. Finally, Section \ref{sec:concl} concludes. The appendix collects various theoretical extensions, additional simulation results, additional material on the application, and all the proofs. The companion R package \href{https://CRAN.R-project.org/package=RationalExp}{RationalExp}, described in the user guide \citep*[][]{RationalExp_R18}, performs the test of RE.

\section{Set-up and characterizations}\label{sec2}

\subsection{Set-up}\label{ss21}

We assume that the researcher has access to a first dataset containing the individual outcome variable of interest, which we denote by $Y$. She also observes, through a second dataset drawn from the same population, the elicited individual expectation on $Y$, denoted by $\psi$. The two datasets, however, cannot be matched. We focus on situations where the researcher has access to elicited beliefs about mean outcomes, as opposed to probabilistic expectations about the full distribution of outcomes. The type of subjective expectations data we consider in the paper has been collected in various contexts, and used in a number of prior studies \citep*[see, among others,][]{Delavande2008,zafar11,AHK2012,AHMR2014,HoffmanBurks17}.

\medskip

\medskip
Formally, $\psi=\mathcal{E}[Y|\mathcal{I}]$, where  $\mathcal{I}$ denotes the $\sigma$-algebra corresponding to the agent's information set and $\mathcal{E}\left[ \cdot | \mathcal{I} \right]$ is the subjective expectation operator (i.e. for any $U$, $\mathcal{E}\left[ U| \mathcal{I} \right]$ is a $\mathcal{I}$-measurable random variable). We are interested in testing the rational expectations (RE) hypothesis $\psi=\E[Y|\mathcal{I}]$, where $ \E\left[\cdot | \mathcal{I}\right] $ is the conditional expectation operator generated by the true data generating process. Importantly, we remain agnostic throughout most of our analysis on the information set $\mathcal{I}$. Our setting is also compatible with heterogeneity in the information different agents use to form their expectations. To see this, let $(U_1,...,U_m)$ denote $m$ variables that agents may or may not observe when they form their expectations, and let $D_k=1$ if $U_k$ is observed, 0 otherwise. Then, if $\mathcal{I}$ is the information set generated by $(D_1 U_1,...,D_mU_m)$, agents will use different subsets of the $(U_k)_{k=1...m}$ (i.e., different pieces of information) depending on the values of the $(D_k)_{k=1...m}$. Our setup encompasses a wide variety of situations, where individuals have private information and form their beliefs based on their information set. This includes various contexts where individuals form their expectations about future outcomes, including education, labor market as well as health outcomes. By remaining agnostic on the information set, our analysis complements several studies which primarily focus on testing for different information sets, while maintaining the rational expectations assumption (see \citeauthor{CunhaHeckman07}, \citeyear{CunhaHeckman07}, for a survey).

\medskip
It is easy to see that the RE hypothesis imposes restrictions on the joint distribution of realizations $Y$ and beliefs $\psi$. In this data combination context, the relevant question of interest is then whether one can rationalize RE, in the sense that there exists a triplet $(Y',\psi',\mathcal{I}')$ such that $(i)$ the pair of random variables $(Y',\psi')$ are compatible with the marginal distributions of $Y$ and $\psi$; and $(ii)$ $\psi'$ correspond to the rational expectations of $Y'$, given the information set $\mathcal{I}'$, i.e., $\E(Y'|\mathcal{I}')=\psi'$. Hence, we consider the test of the following hypothesis:
\begin{align*}
\text{H}_0: \ & \text{there exists a pair of random variables } (Y',\psi')  \text{ and a sigma-algebra } \mathcal{I}' \text{ such that } \\
&  \sigma(\psi') \subset \mathcal{I}', Y' \sim Y, \; \psi' \sim \psi  \text{ and } \E\left[Y'\middle|\mathcal{I}'\right]=\psi',
\end{align*}
where $\sim$ denotes equality in distribution. Rationalizing RE does not mean that the true realizations $Y$, beliefs $\psi$ and information set $\mathcal{I}$ are such that $\E\left[Y|\mathcal{I}\right]=\psi$. Instead, it means that there exists a triplet $\left(Y',\psi', \mathcal{I}'\right)$ consistent with the data and such that $\E\left[Y'|\mathcal{I}'\right]=\psi'$. In other words, rejecting $\text{H}_0$ implies that RE does not hold, in the sense that the true realizations, beliefs, and information set do not satisfy RE ($ \E\left[Y|\mathcal{I}\right]\neq \psi$). The converse, however, is not true.

\subsection{Equivalences}\label{ss22}

\subsubsection{Main equivalence}\label{ss221}

Let $F_{\psi}$ and $F_{Y}$ denote the cumulative distribution functions (cdf) of $\psi$ and $Y$, $x^+ = \max(0, x)$, and define
 $$\Delta(y)= \int_{-\infty}^y   F_{Y} (t) - F_{\psi} (t) dt.$$
Throughout most of our analysis, we impose the following regularity conditions on the distributions of realized outcomes ($Y$) and subjective beliefs ($\psi$):

\begin{assumption}\label{hyp:regul_equiv}
	$\E\left(|Y|\right)<\infty$ and $\E\left(|\psi|\right)<\infty$.
 \end{assumption} \vspace{1mm}

The following preliminary result will be useful subsequently.

\begin{lemma}\label{lem:reformulation}
	Suppose that Assumption \ref{hyp:regul_equiv} holds. Then H$_0$ holds if and only if there exists a pair of random variables $(Y',\psi')$ such that  $Y' \sim Y$, $\psi' \sim \psi$ and $\E\left[Y'\middle|\psi'\right]=\psi'$.
\end{lemma}

Lemma \ref{lem:reformulation} states that in order to test for H$_0$, we can focus on the constraints on the joint distribution of $Y$ and $\psi$, and ignore those related to the information set. This is intuitive given that we impose no restrictions on this set. Our main result is Theorem~\ref{prop:equiv} below. It states that rationalizing RE (i.e., $\text{H}_0$) is equivalent to a continuum of moment inequalities,  and one moment equality.

\begin{theorem}\label{prop:equiv}
Suppose that Assumption~\ref{hyp:regul_equiv} holds. The following statements are equivalent:
\begin{enumerate}
\item[(i)] $\text{H}_0$ holds;
\item[(ii)] ($F_Y$ mean-preserving spread of $F_\psi$) $\Delta(y)\geq 0$ for all $y\in \R$ and $\E\left[Y\right]=\E\left[\psi\right]$;
\item[(iii)]$\E\left[    \left(y-Y\right)^+ -  \left(y-\psi\right)^+\right] \geq 0  $ for all $y\in \R$  and  $\E\left[Y\right]=\E\left[\psi\right]$.
\end{enumerate}
\end{theorem}\medskip

The implication (i) $\Rightarrow$ (iii) and the equivalence between (ii) and (iii) are simple to establish. The key part of the result is to prove that (iii) implies (i). To show this, we first use Lemma~\ref{lem:reformulation}, which states that H$_0$ is equivalent to the existence of $(Y',\psi')$ such that $Y'\sim Y$, $\psi'\sim \psi$ and $\E\left[Y'\middle|\psi'\right]=\psi'$. Then the result essentially follows from Strassen's theorem \citep[][Theorem 8]{strassen1965}.\medskip

It is interesting to note that Theorem \ref{prop:equiv} is related to the theory of risk in microeconomic theory. In particular, using the terminology of \cite{rothschild1970increasing}, (ii) states that realizations ($Y$) are more risky than beliefs ($\psi$). The main value of Theorem \ref{prop:equiv}, from a statistical point of view, is to transform $\text{H}_0$ into the set of moment inequality (and equality) restrictions given by (iii). We show in Section \ref{sub:stat_tests} how to build a statistical test of these conditions.

\medskip
\paragraph{Comparison with alternative approaches}

We now compare our approach with alternative ones that have been proposed in the literature. In the following discussion, as in this whole section, we reason at the population level and thus ignore statistical uncertainty. Accordingly, the ``tests'' we consider here are formally deterministic, and we compare them in terms of data generating processes violating the null hypothesis associated with each of them.

\medskip
Our approach can clearly detect many more violations of rational expectations than the ``naive'' approach based solely on the equality $\E(Y)=\E(\psi)$. It also detects more violations than the approach based on the restrictions $\E(Y)=\E(\psi)$ and $\V(Y)\geq\V(\psi)$ (approach based on the variance), which has been considered in particular in the macroeconomic literature on the accuracy and rationality of forecasts \citep[see, e.g.][]{patton2012forecast}. On the other hand, and as expected since it relies on the joint distribution of $(Y,\psi)$, the ``direct'' approach for testing RE, based on $\E(Y|\psi)=\psi$, can detect more violations of rational expectations than ours.

\medskip
To better understand the differences between these four different approaches (``naive'', variance, ``direct'', and ours), it is helpful to consider important particular cases. Of course, if $\psi=\RE$, individuals are rational and none of the four approaches leads to reject RE. Next, consider departures from rational expectations of the form $\psi=\RE+\eta$, with $\eta$ independent of $\RE$. If $\E(\eta)\neq 0$, subjective beliefs are biased, and individuals are on average either over-pessimistic or over-optimistic. It follows that $\E(Y)\neq \E(\psi)$, implying that all four approaches lead to reject RE.

\medskip
More interestingly, if $\E(\eta)=0$, individuals' expectations are right on average, and the naive approach does not lead to reject RE. However, it is easy to show that, as long as deviations from RE are heterogeneous in the population ($\mathbb{V}(\eta)>0$), the direct approach always leads to a rejection. In this setting, our approach constitutes a middle ground, in which rejection of RE depends on the degree of dispersion of the deviations from RE ($\eta$) relative to the uncertainty shocks ($\eps=Y - \E(Y|\mathcal{I})$). In other words and intuitively, we reject RE whenever departures from RE dominate the uncertainty shocks affecting the outcome. Formally, and using similar arguments as in Proposition \ref{prop:meas_err} in Subsection~\ref{sub:meas_err}, one can show that if $\eps$ is independent of $\RE$, we reject $\text{H}_0$ as long as the distribution of the uncertainty shocks stochastically dominates at the second-order the distribution of the deviations from RE.

\medskip
Specifically, if $\eps \sim \mathcal{N}(0,\sigma^2_\eps)$ and $\eta \sim \mathcal{N}(0,\sigma^2_\eta)$, we reject RE if and only if $\sigma^2_\eta > \sigma^2_\eps$. In such a case, our approach boils down to the variance approach mentioned above: we reject whenever $\mathbb{V}(\psi)>\mathbb{V}(Y)$. But interestingly, if the discrepancy ($\eta$) between beliefs and RE is not normally distributed, we can reject H$_0$ even if $\mathbb{V}(\psi) \leq \mathbb{V}(Y)$. Suppose for instance that $\eps\sim\mathcal{N}(0,1)$ and
$$\eta = a \left(-\mathds{1}\{U \leq 0.1\} + \mathds{1}\{U \geq 0.9\}\right), \; U \sim \mathcal{U}[0,1] \text{ and } a>0.$$
In other words, 80\% of individuals are rational, 10\% are over-pessimistic and form expectations equal to $\RE-a$, whereas 10\% are over-optimistic and expect $\RE + a$. Then one can show that our approach leads to reject RE when $a\geq 1.755$, while for $a= 1.755$, $\mathbb{V}(\eta)\simeq 0.616 < \mathbb{V}(\eps)=1$.

\medskip
\paragraph{Binary outcome}

Our equivalence result does not require the outcome $Y$ to be continuously distributed. In the particular case where $Y$ is binary, our test reduces to the naive test of $\E(Y)=\E(\psi)$. Indeed, when $Y$ is a binary outcome and $\psi\in[0,1]$, one can easily show that as long as $\E(Y)=\E(\psi)$, the inequalities $\E\left[    \left(y-Y\right)^+ -  \left(y-\psi\right)^+\right] \geq 0$ automatically hold for all $y\in \R$. This applies to expectations about binary events, such as, e.g., being employed or not at a given date.

\medskip
\paragraph{Interpretation of the boundary condition} 
To shed further light on our test and on the interpretation of $\text{H}_0$, it is instructive to derive the distributions of $Y|\psi$ that correspond to the boundary condition ($\Delta(y)=0$). The proposition below shows that, in the presence of rational expectations, agents whose beliefs $\psi$ lies at the boundary of $\text{H}_0$ have perfect foresight, i.e. $\psi =\E[Y|\mathcal{I}]= Y$.\footnote{For any cdf $F$, we let $F^{-1}$ denote its quantile function, namely $F^{-1}(\tau)=\inf\{x:F(x)\geq \tau\}$.}

\begin{proposition}\label{prop:1}
	Suppose that $(Y,\psi)$ satisfies RE, $u\mapsto F^{-1}_{Y|\psi}(\tau|u)$ is continuous for all $\tau \in(0,1)$, and $\Delta(y_0)=0$ for some $y_0$ in the interior of the support of $\psi$. Then the distribution of $Y$ conditional on $\psi=y_0$ is degenerate: $P(Y=y_0|\psi=y_0)=1$. 
\end{proposition}

\subsubsection{Equivalence with covariates}\label{ss222}

In practice we may observe additional variables $X\in \R^{d_X}$ in both datasets. Assuming that $X$ is in the agent's information set, we modify H$_0$ as follows:\footnote{See complementary work by \cite{GHP16}, who use subjective expectations data to relax the rational expectations assumption, and propose a method allowing to test whether specific covariates are included in the agents' information sets.}
\begin{align*}
\text{H}_{0X}: \ & \text{there exists a pair of random variables } (Y',\psi')  \text{ and a sigma-algebra } \mathcal{I}' \text{ such that } \\
& \sigma(\psi', X) \subset \mathcal{I}', \, Y'|X\sim Y|X , \, \psi'|X \sim \psi|X \text{ and } \E\left[Y'\middle|\mathcal{I}'\right]=\psi'.
\end{align*}

Adding covariates increases the number of restrictions that are implied by the rational expectation hypothesis, thus improving our ability to detect violations of rational expectations. Proposition \ref{prop:equivX} below formalizes this idea and shows that H$_{0X}$ can be expressed as a continuum of conditional moment inequalities, and one conditional moment equality.

\begin{proposition}\label{prop:equivX}
Suppose that Assumption~\ref{hyp:regul_equiv} holds. The following two statements are equivalent:
\begin{enumerate}
\item[(i)] $\text{H}_{0X}$ holds;
\item[(ii)] Almost surely, $  \E\left[ \left(y-Y\right)^+ - \left(y-\psi\right)^+  \middle| X \right] \geq 0  $ for all $y\in \R$  and  $ \E\left[  Y - \psi\middle| X\right] =0 $.
\end{enumerate}
Moreover, if $\text{H}_{0X}$ holds, $\text{H}_0$ holds as well.
\end{proposition}

\subsubsection{Equivalence with unpredictable aggregate shocks}\label{ss223}

Oftentimes, the outcome variable is affected not only by individual-specific shocks, but also by aggregate shocks. We denote by $C$ the random variable corresponding to the aggregate shocks. The issue, in this case, is that we observe a single realization of $C$ ($c$, say), along with the outcome variable conditional on that realization $C=c$. In other words, we only identify $F_{Y|C=c}$ rather than $F_Y$, as the latter would require to integrate over the distribution of all possible aggregate shocks. Moreover, the restriction $\E\left[Y\middle|C=c, \psi\right]= \psi$ is generally violated, even though the rational expectations hypothesis holds. It follows that one cannot directly apply our previous results by simply replacing $F_Y$ by $F_{Y|C=c}$. In such a case, one has to make additional assumptions on how the aggregate shocks affect the outcome.

\medskip
To illustrate our approach, let us consider the example of individual income. Suppose that the logarithm of income of individual $i$ at period $t$, denoted by $Y_{it}$, satisfies a Restricted Income Profile model:
$$Y_{it} = \alpha_i + \beta_t + \eps_{it},$$
where $\beta_t$ capture aggregate (macroeconomic) shocks, $\eps_{it}$ follows a zero-mean random walk, and $\alpha_i$, $(\beta_{t})_{t}$ and $(\eps_{it})_t$ are assumed to be mutually independent. Let $\mathcal{I}_{it-1}$ denote individual $i$'s information set at time $t-1$, and suppose that $\mathcal{I}_{it-1}=\sigma\left(\alpha_i, (\beta_{t-k})_{k\geq 1},
(\eps_{it-k})_{k\geq 1}\right)$. If individuals form rational expectations on their future outcomes, their beliefs in period $t-1$ about their future log-income in period $t$ are given by
$$\psi_{it} = \E\left[Y_{it}\middle|\mathcal{I}_{it-1}\right] = \alpha_i + \E\left[\beta_t\middle|(\beta_{t-k})_{k\geq 1}\right] + \eps_{it-1}.$$
Thus, $Y_{it} = \psi_{it} + C_t + \eps_{it}-\eps_{it-1}$, with $C_t =\beta_t - \E\left[\beta_t\middle|(\beta_{t-k})_{k\geq 1}\right]$. The corresponding conditional expectation is given by:
$$\E\left[Y_{it} \middle|\mathcal{I}_{it-1},C_t=c_t\right]=\psi_{it} + c_t \neq \psi_{it}.$$

\medskip
To get closer to our initial set-up, we now drop indexes $i$ and $t$ and maintain the conditioning on the aggregate shocks $C=c$ implicit. Under these conventions, rationalizing RE does not correspond to $\RE=\psi$, but instead to $\RE=c_0 + \psi$ for some $c_0\in\R$. A similar reasoning applies to multiplicative instead of additive aggregate shocks. In such a case, the null takes the form $\RE=c_0\psi$, for some $c_0>0$. In these two examples, $c_0$ is identifiable: by $c_0 = \E(Y) - \E(\psi)$ in the additive case, by $c_0=\E(Y)/\E(\psi)$ in the multiplicative case. Moreover, there exists in both cases a known function $q(y,c)$ such that $\E(q(Y,c_0))=\E(\psi)$, namely $q(y,c)=y-c$ and $q(y,c)=y/c$ for additive and multiplicative shocks, respectively.

\medskip
More generally, we consider the following null hypothesis for testing RE in the presence of aggregate shocks:
\begin{align*}
\text{H}_{0S}:& \; \text{there exist random variables }  \left(Y', \psi'\right), \text{ a sigma-algebra } \mathcal{I}' \text{ and } c_0\in\R \text{ such that } \\
 & \sigma(\psi') \subset \mathcal{I}', \, Y' \sim Y, \, \psi'\sim \psi \text{ and } \E\left[q\left(Y',c_0\right)\middle|\mathcal{I}'\right]=\psi'.
\end{align*}
where $q(.,.)$ is a known function supposed to satisfy the following restrictions.

\begin{assumption}
	\label{hyp:for_test_gen} $\E\left(|\psi|\right)<\infty$ and for all $c$, $\E\left(|q\left(Y,c\right)|\right)<\infty$. Moreover, $\E\left[q(Y,c)\right]=\E[\psi]$ admits a unique solution, $c_0$.
\end{assumption}

By applying our main equivalence result (Theorem \ref{prop:equiv}) to $q(Y,c_0)$ and $\psi$, we obtain the following result.

\begin{proposition}\label{prop:equiv_gen}
Suppose that Assumption \ref{hyp:for_test_gen} holds. Then the following statements are equivalent:
\begin{enumerate}
\item[(i)] $\text{H}_{0S}$ holds;
\item[(ii)] $  \E\left[    \left(y- q\left(Y,c_0\right) \right)^+ -  \left(y-\psi\right)^+ \right] \geq 0  $ for all $y\in \R$.
\end{enumerate}
\end{proposition}

A few remarks on this proposition are in order. First, this result can be extended in a straightforward way to a setting with covariates. This is important not only to increase the ability of our test to detect violations of RE, but also because this allows for aggregate shocks that differ across observable groups. We discuss further this extension, and the corresponding statistical test, in Appendix \ref{app:test_shock}. Second, in the presence of aggregate shocks, the null hypothesis does not involve a moment equality restriction anymore; the corresponding moment is used instead to identify $c_0$. Related, a clear limitation of the naive test ($\E(Y)=\E(\psi)$) is that, unlike our test, it is not robust to aggregate shocks. In this case, rejecting the null could either stem from violations of the rational expectation hypothesis, or simply from the presence of aggregate shocks. Third, in Appendix~\ref{subsec:imp_shock}, we examine whether one can extend the results above to test for RE when aggregate shocks affect the outcomes in a more general way. Proposition~\ref{prop:imposs_result} establishes a negative result in this respect: as long as one allows for a sufficiently flexible dependence between the outcome and the aggregate shocks, any given distribution of subjective expectations is arbitrarily close to a distribution for which RE can be rationalized. This implies that, within this more general class of outcome models, there does not exist any almost-surely continuous RE test that has non-trivial power. 

\subsubsection{Robustness to measurement errors} % (fold)
\label{sub:meas_err}

We have assumed so far that $Y$ and $\psi$ were perfectly observed; yet measurement errors in survey data are pervasive \citep*[see, e.g.][]{BBM2001}. We explore in the following the extent to which our test is robust to measurement errors. By robust, we mean that the test does not incorrectly reject RE, when they in fact hold. Specifically, assume that the true variables ($\psi$ and $Y)$ are unobserved. Instead, we only observe $\widehat{\psi}$ and $\widehat{Y}$, which are affected by classical measurement errors.\footnote{See \cite{zafar2011can} who does not find evidence of non-classical measurement errors on subjective beliefs elicited from a sample of Northwestern undergraduate students. We conjecture that our test is robust to some forms of non-classical measurement errors. However, it seems difficult in this case to obtain a general result similar to the one in Proposition \ref{prop:meas_err}.} Namely:
\begin{equation}
\begin{array}{rcl}
\widehat{\psi} = \psi + \xi_\psi & \text{with} &  \xi_\psi \indep \psi, \; \E[\xi_\psi]=0 \\[2mm]
\widehat{Y} =Y+ \xi_Y & \text{with} & \xi_Y\indep Y, \; \E[\xi_Y]=0. 	
\end{array}
\label{eq:meas_errors}
\end{equation}
The following proposition shows that our test is robust to a certain degree of measurement errors on the beliefs. 

\begin{proposition}\label{prop:meas_err}
Suppose that $Y$ and $\psi$ satisfy H$_0$, and let $\eps=Y-\psi$ and $\left(\widehat{\psi}, \widehat{Y}\right)$ be defined as in \eqref{eq:meas_errors}. Suppose also that $\eps+\xi_Y \indep \psi$ and $F_{\xi_\psi}$ dominates at the second order $F_{\xi_Y+\eps}$. Then $\widehat{Y}$ and $\widehat{\psi}$ satisfy H$_0$.
\end{proposition}

\medskip
The key condition is that $F_{\xi_\psi}$ dominates at the second order $F_{\xi_Y+\eps}$, or, equivalently here, that $F_{\xi_Y+\eps}$ is a mean-preserving spread of $F_{\xi_\psi}$. Recall that in the case of normal variables, $\xi_\psi \sim \mathcal{N}(0,\sigma^2_1)$ and $\xi_Y+\eps \sim \mathcal{N}(0,\sigma^2_2)$, this is in turn equivalent to imposing $\sigma^2_1 \leq \sigma^2_2$. Thus, even if there is no measurement error on $Y$, so that $\xi_Y=0$, this condition may hold provided that the variance of measurement errors on $\psi$ is smaller than the variance of the uncertainty shocks on $Y$. More generally, this allows elicited beliefs to be - potentially much - noisier than realized outcomes, a setting which is likely to be relevant in practice. One should not infer, however, that measurement errors are innocuous in our set-up. Indeed, the converse of Proposition \ref{prop:meas_err} does not hold: we may reject $H_0$ with $Y$ and $\psi$, but not with $\widehat{Y}$ and $\widehat{\psi}$. As a simple example, suppose that $Y\sim\mathcal{N}(0,\sigma^2_Y)$, $\psi\sim\mathcal{N}(0,\sigma^2_\psi)$, $\xi_Y\sim \mathcal{N}(0,\sigma^2_3)$, $\xi_\psi=0$ and $\sigma^2_\psi \in (\sigma^2_Y, \sigma^2_Y+\sigma^2_3]$. Then, $\widehat{Y}$ and $\widehat{\psi}$ satisfy $H_0$, since $\sigma^2_\psi \leq \sigma^2_Y+\sigma^2_3$, whereas $Y$ and $\psi$ do not, since $\sigma^2_\psi >\sigma^2_Y$. Importantly though, Proposition \ref{prop:meas_err}  does show that our test is conservative in the sense that measurement errors cannot result in incorrectly concluding that the RE hypothesis does not hold.

\medskip
In situations where $(\widehat{Y},\widehat{\psi})$ are jointly observed, one could in principle alternatively implement the direct test. However, in contrast to our test, %that is based on the marginal distributions, 
the direct test is not robust to any measurement errors on the subjective beliefs $\psi$. Indeed, if RE holds, so that $\E\left[Y|\psi\right]=\psi$, it is nevertheless the case that $\E\left[\widehat{Y}\middle|\widehat{\psi}\right]\neq \widehat{\psi}$, as long as $\Cov(\xi_Y,\widehat{\psi})=\Cov(\xi_\psi, Y)=0$ and $\mathbb{V}(\xi_\psi)>0$. In other words, even if individuals have rational expectations, the direct test will reject the null hypothesis in the presence of even an arbitrarily small degree of measurement errors on the elicited beliefs.

\medskip
Also, it is unclear whether, in the presence of measurement errors on the elicited beliefs and beyond the restrictions on the marginal distributions, there are restrictions on the copula of $(\widehat{Y},\widehat{\psi})$ that are implied by RE. For instance, we show in Proposition \ref{prop:reg_meas_err} in Appendix \ref{sec:lin_reg_meas_err} that under RE, and without imposing restrictions on the dependence between  $\xi_Y+\eps$ and $\xi_\psi$, the coefficient of the (theoretical) linear regression of $\widehat{Y}$ on $\widehat{\psi}$ remains unrestricted.\footnote{There might of course possibly be additional relevant information in the higher-order moments, although we have not been able to find any.} On the other hand, if one assumes that  $\Cov(\xi_Y+\eps, \xi_\psi)\geq 0$ and $\V(\psi)/\V(\xi_\psi)\geq \underline{\lambda}$ for some $\underline{\lambda}\geq 0$, Proposition \ref{prop:reg_meas_err} also shows that the coefficient of the linear regression of $\widehat{Y}$ on $\widehat{\psi}$ is bounded from below under RE. Such a restriction, which does require to take a stand on the signal-to-noise ratio $\V(\psi)/\V(\xi_\psi)$, can be easily added to the moment inequalities of our test if $(\widehat{Y},\widehat{\psi})$ is observed.

\subsubsection{Other extensions} % (fold)
\label{sub:extensions}

We now briefly discuss other relevant directions in which Theorem \ref{prop:equiv} can be extended. First, another potential source of uncertainty on $\psi$ is rounding. Rounding practices by interviewees are common in the case of subjective beliefs. Under additional restrictions, it is possible in such a case to construct bounds on the true beliefs $\psi$ \citep[see, e.g.,][]{manski2010rounding}. We show in Appendix \ref{sec:test_bounds} that our test can be generalized to accommodate this rounding practice.

\medskip
Second, we have implicitly maintained the assumption so far that subjective beliefs and realized outcomes are drawn from the same population. In Appendix \ref{sec:propensity_score}, we relax this assumption and show that our test can be easily extended to allow for sample selection under unconfoundedness, through an appropriate reweighting of the observations.

\medskip
Third, our equivalence result and our test can be extended to accommodate situations with multiple outcomes $(Y_k)_{k=1,..,K}$ and multiple subjective beliefs $(\psi_k)_{k=1,..,K}$ associated with each of these outcomes. Specifically, whether one can rationalize rational expectations in this environment can be written as:
\begin{equation*}
\E(Y_k|\psi_1,...,\psi_K)=\psi_k \textrm{, for all } k \in \{1,...,K\}
\end{equation*}
which, in turn, is equivalent to the distribution of the outcomes $Y_k$ being a mean-preserving spread of the distribution of the beliefs $\psi_k$. This situation arises in various contexts, including cases where respondents declare their subjective probabilities of making particular choices among $K+1$ possible alternatives. This also arises in situations where expectations about the distribution of a continuous outcome $Y$ are elicited through questions of the form ``what do you think is the percent chance that [Y] will be greater than [y]?'', for different values $(y_k)_{k=1,..,K}$.  In such cases, it is natural to build a RE test based on the multiple outcomes $(\mathds{1}\{Y> y_k\})_{k=1,..,K}$ and subjective beliefs $(\psi_k)_{k=1,..,K}$, where $\psi_k$ is the subjective survival function of $Y$ evaluated at $y_k$.

% subsubsection other_extensions (end)

\section{Statistical tests}
\label{sub:stat_tests}

We now propose a testing procedure for H$_{0X}$, which can be easily adapted to the case where no covariate common to both datasets is available to the analyst. To simplify notation, we use a potential outcome framework to describe our data combination problem. Specifically, instead of observing $(Y,\psi)$, we suppose to observe only, in addition to the covariates $X$, $\widetilde{Y} = DY + (1-D)\psi $ and $D$, where $D=1$ (resp. $D=0$) if the unit belongs to the dataset of $Y$ (resp. $\psi$). As in Subsection~\ref{ss21}, we assume that the two samples are drawn from the same population, which amounts to supposing that $D\indep (X, Y,\psi)$ (see Assumption \ref{hyp:regul}-(i) below). In order to build our test, we use the characterization (ii) of Proposition \ref{prop:equivX}:
$$  \E\left[ \left(y-Y\right)^+  - \left(y-\psi\right)^+\middle| X \right] \geq 0 \quad \forall y\in \R \; \text{ and}  \quad \E\left[  Y - \psi\middle| X\right] =0 .$$
Equivalently but written more compactly with $\widetilde{Y}$ only,
$$\E\left[ W \left(y-\widetilde{Y}\right)^+\middle|X\right]\geq 0  \quad \forall y\in \R \; \text{ and} \quad \E\left[W\widetilde{Y}\middle|X\right]= 0,$$
where $W=D/\E(D)-(1-D)/\E(1-D)$. This formulation of the null hypothesis allows us to apply the instrumental functions approach of \citeauthor{andrews2016inference} (\citeyear{andrews2016inference}, AS), who consider the issue of testing many conditional moment inequalities and equalities. We then build on their results to establish that our test controls size asymptotically and is consistent over fixed alternatives.\footnote{Other testing procedures could be used to implement our test, such as that proposed by \cite{linton2010improved}.} The initial step is to transform the conditional moments into the following unconditional moments conditions:
$$\E\left[W\left(y-\widetilde{Y}\right)^+ g(X)\right] \geq 0,  \quad \E\left[\left(  Y - \psi\right)g(X)\right] =0 ,$$
for all $y\in \R$ and $g$ belonging to a suitable class of non-negative functions.

\medskip
We suppose to observe a sample $(D_i, X_i, \widetilde{Y}_i)_{i=1...n}$ of $n$ i.i.d. copies of $(D,X,\widetilde{Y})$. We consider instrumental functions $g$ that are indicators of belonging to specific hypercubes within $[0,1]^{d_X}$, hence we tranform the variables $X_i$ to lie in $[0,1]^{d_X}$. For notational convenience, we let $\widetilde{X}_i$ denote the nontransformed vector of covariates, and redefine $X_i$ as:
$$X_i=\Phi_0\left( \widehat{\Sigma}_{ \widetilde{X},n}^{-1/2}\left(  \widetilde{X}_{i} - \overline{ \widetilde{X}}_{i}\right)\right),$$
where, for any $x=(x_1,\dots,x_{d_X})$, we let $\Phi_0(x)=\left( \Phi(x_1) , \dots, \Phi\left(x_{d_X} \right)\right)^{\top}$. Here $ \Phi $ denotes the standard normal cdf,  $  \widehat{\Sigma}_{\widetilde{X},n} $ is the sample covariance matrix of $ \left(\widetilde{X}_i\right)_{i=1...n}$ and  $ \overline{\widetilde{X}}_n $ its sample mean.

\medskip
Specifically, we consider instrumental functions $g$ belonging to the class of functions $\mathcal{G}_r= \left\{g_{a,r}, \; a\in A_r\right\}$, with
$A_r= \left\{1,2,\dots, 2r\right\}^{d_X}$ ($r\geq 1$), $g_{a,r}(x) = \indic\left\{x \in  C_{a,r}\right\}$ and, for any $a=(a_1,...,a_{d_X})^{\top}\in A_r$,
$$C_{a,r} = \prod_{u=1}^{d_X} \left(\dfrac{a_u -1 }{2r}, \dfrac{a_u }{2r}\right].$$

\medskip
Finally, to define the test statistic $T$, we need to introduce additional notations. First, let $w_i=nD_i/\sum_{j=1}^n D_j - n(1-D_i)/\sum_{j=1}^n( 1-D_j)$ and define, for any $y\in\R$,
	\begin{eqnarray}\label{eq:mn}
 m\left(D_i,\widetilde{Y}_i, X_i, g,y\right) =\left( \begin{array}{c}  m_{1}\left(D_i,\widetilde{Y}_i, X_i, g,y\right) \\  m_{2}\left(D_i,\widetilde{Y}_i, X_i, g,y\right)
 	\end{array} \right)
	= \left( \begin{array}{c} w_i\left(y - \widetilde{Y}_i\right)^+g\left(X_i\right) \\
	 w_i\widetilde{Y}_i g\left(X_i\right)
	\end{array} \right).
\end{eqnarray}
Let $\overline{m}_n (g,y)= \sum_{i=1}^{n} m\left(D_i,\widetilde{Y}_i, X_i, g,y\right)/n $ and define similarly $\overline{m}_{n,j}$ for $j=1,2$. For any function $g$ and any $y\in\R$, we also define, for some $\epsilon>0$,
 $$ \overline{\Sigma}_n(g,y) = \widehat{\Sigma}_n(g,y) + \epsilon \mathrm{Diag}\left(   \widehat{\mathbb{V}}\left(\widetilde{Y} \right)  , \widehat{\mathbb{V}}\left(\widetilde{Y} \right)     \right),$$
  where $\widehat{\Sigma}_n(g,y) $ is the sample covariance matrix of $ \sqrt{n}\overline{m}_n\left( g,y\right) $ and
  $\widehat{\mathbb{V}}\left(\widetilde{Y} \right)$ is the empirical variance of $\widetilde{Y}$. We then denote by $\overline{\Sigma}_{n,jj}(g,y) \ (j=1,2)$ the $j$-th diagonal term of $\overline{\Sigma}_{n}(g,y)$.

  \medskip
  Then the (Cram\'{e}r-von-Mises) test statistic $T$ is defined by
\begin{align*}
T= &\underset{y \in \widehat{\mathcal{Y}}}{\sup} \sum_{r=1}^{r_n}\dfrac{(2r)^{-d_X}}{\left(r^2 + 100\right) } \sum_{a\in A_r} \bigg[ \left(1-p\right)\left(- \df{\sqrt{n} \overline{m}_{n,1}\left(g_{a,r},y\right)}{\overline{\Sigma}_{n,11}(g_{a,r},y)^{1/2} } \right)^{+ 2} + p \left(  \df{\sqrt{n} \overline{m}_{n,2}\left(g_{a,r},y\right)}{\overline{\Sigma}_{n,22}(g_{a,r},y)^{1/2}}  \right)^2\bigg],
\end{align*}
where $ \widehat{\mathcal{Y}} =\left[\underset{i=1,\dots,n}{\min}\widetilde{Y}_i,\underset{i=1,\dots,n}{\max}\widetilde{Y}_i \right] $, $p\in (0,1)$ is a parameter weighting the moments inequalities versus equalities and $(r_n)_{n\in\N}$ is a deterministic sequence tending to infinity.

\medskip
To test for rational expectations in the absence of covariates, we set the instrumental function equal to the constant function $g(X)=1$, and the test statistic is simply written as:
$$ T= \underset{y \in \widehat{\mathcal{Y}}}{\sup}\bigg[  \left(1-p\right)\left( - \df{\sqrt{n} \overline{m}_{n,1}(y)}{\overline{\Sigma}_{n,11}(y)^{1/2} } \right)^{+2} + p \left(  \df{\sqrt{n} \overline{m}_{n,2}(y)}{\overline{\Sigma}_{n,22}(y)^{1/2}}  \right)^2\bigg],$$
where, using the notations introduced above, $\overline{m}_{n,j}(y)=\overline{m}_{n,j}(1,y)$ and $\overline{\Sigma}_{n,jj}(y)=\overline{\Sigma}_{n,jj}(1,y)$ $(j=1,2)$.

\medskip
Whether or not covariates are included, the resulting test is of the form $\varphi_{n,\alpha} = \indic\left\{T> c^*_{n,\alpha}\right\}$ where the estimated critical value $c^*_{n,\alpha}$ is obtained by bootstrap using as in AS the Generalized Moment Selection method. Specifically, we follow three steps:
\begin{enumerate}
\item Compute the function $ \overline{\varphi}_n\left(y,g\right) = \left(\overline{\varphi}_{n,1}\left(y,g\right), 0  \right)^{\top} $ for $ (y,g) $ in $ \widehat{\mathcal{Y}}\times\cup_{r=1}^{r_n} \mathcal{G}_{r} $, with
\[  \overline{\varphi}_{n,1}\left(y,g\right) = \overline{\Sigma}_{n,11}^{1/2} B_n\indic\left\{   \df{n^{1/2}}{\kappa_n} \overline{\Sigma}_{n,11}^{-1/2}\overline{m}_{n,1}(y,g)  >1  \right\} , \]
and where $ B_n =  \left(b_0\ln(n)/\ln(\ln(n))\right)^{1/2} $, $ b_0 >0 $, $ \kappa_n =(\kappa\ln(n))^{1/2}$, and $\kappa>0$. To compute $\overline{\Sigma}_{n,11}$, we fix $\epsilon$ to $0.05$, as in AS.
\item Let $\left(D_i^*,\widetilde{Y}_i^*, X_i^*\right)_{i=1,...,n}$ denote a bootstrap sample, i.e., an i.i.d. sample from the empirical cdf of $\left(D,\widetilde{Y}, X\right)$, and compute from this sample the bootstrap counterparts of $ \overline{m}_n$ and $ \overline{\Sigma}_{n}$,  $ \overline{m}_n^* $ and $ \overline{\Sigma}_{n}^* $. Then compute the bootstrap counterpart of $T$, $T^*$, replacing $ \overline{\Sigma}_{n}\left(y,g_{a,r}\right)$  and $ \sqrt{n}\overline{m}_n\left(y,g_{a,r}\right) $ by  $ \overline{\Sigma}_{n}^*\left(y,g_{a,r}\right)$ and $\sqrt{n}\left( \overline{m}_n^* - \overline{m}_n \right)\left(y,g_{a,r}\right) +  \overline{\varphi}_n\left(y,g_{a,r}\right)$, respectively.
\item The threshold $c^*_{n,\alpha}$ is the quantile (conditional on the data) of order $1-\alpha + \eta$ of $T^* +\eta$ for some $ \eta >0 $. Following AS, we set $\eta$ to $10^{-6} $.
\end{enumerate}

Note that, despite the multiple steps involved, the testing procedure remains computationally easily tractable. In particular, for the baseline sample we use in our application (see Section~\ref{subsec:data}), the RE test only takes 2 minutes.\footnote{This CPU time is obtained using our companion R package, on an Intel Xeon CPU E5-2643, 3.30GHz with 256Gb of RAM.}

\medskip
We now turn to the asymptotic properties of the test. For that purpose, it is convenient to introduce additional notations. Let $\mathcal{Y}$ and $\mathcal{X}$ denote the support of $Y$ and $X$ respectively, and
$$\mathcal{L}_F= \left\{ \left(y,g_{a,r}\right): y\in \mathcal{Y}, \ (a,r)\in A_r\times\N: \ \E_{F}\left[ W \left(y -\widetilde{Y}\right)^+ g_{a,r}(X)\right]=0 \right\},$$
where, to make the dependence on the underlying probability measure explicit, $\E_F$ denotes the expectation with respect to the distribution $F$ of $\left(D,\widetilde{Y},X\right)$.
Finally, let $\mathcal{F}$ denote a subset of all possible cumulative distribution functions of $\left(D,\widetilde{Y},X\right)$ and $\mathcal{F}_0$ be the subset of $\mathcal{F}$ such that H$_{0X}$ holds. We impose the following conditions on $\mathcal{F}$ and $\mathcal{F}_0$. 

\medskip
\begin{assumption}\label{hyp:regul}~\\
	\vspace{-0.5cm}
\begin{enumerate}
	\item[(i)] For all $F\in \mathcal{F}$, $D\indep (X, Y,\psi)$;
	\item[(ii)] There exists $M>0$ such that $\widetilde{Y}\in [-M,M]$ for all $F\in \mathcal{F}$. Also, $\inf_{F\in \mathcal{F}}\mathbb{V}_F\left(\widetilde{Y}\right)>0$ and $0 < \inf_{F\in \mathcal{F}} \E_F\left[ D \right] \leq \sup_{F\in \mathcal{F}}  \E_F\left[ D \right] < 1$;
	\item[(iii)] For all  $ F\in \F_0$,  $K_{F}$, the asymptotic covariance kernel of $ n^{-1/2} \mathrm{Diag}\left(\mathbb{V}_F\left(\widetilde{Y}\right)\right)^{-1/2} \overline{m}_n$ is in a compact set  $\mathcal{K}_2$ of the set of all $ 2\times2 $ matrix valued covariance kernels on $ \mathcal{Y}\times \cup_{r\geq 1} \mathcal{G}_{r}$ with uniform metric $d$ defined by
	\[ d(K,K')= \underset{(y,g,y',g')\in \left(\mathcal{Y} \times \cup_{r\geq 1} H_{r}\right)^2 }{\sup} \left\|K(y,g,y',g') -  K'(y,g,y',g')\right\|.\]
\end{enumerate}
 \end{assumption}

The main result of this section is Theorem \ref{prop:asympt_size}. It shows that, under Assumption \ref{hyp:regul}, the test $\varphi_{n,\alpha}$ controls the asymptotic size and is consistent over fixed alternatives.

\begin{theorem}\label{prop:asympt_size}
Suppose that $r_n\rightarrow \infty$ and Assumption \ref{hyp:regul} holds. Then:
\begin{enumerate}
	\item[(i)] $\limsup_{n\rightarrow\infty} \sup_{F\in \mathcal{F}_0} 
\E_F[\varphi_{n,\alpha}]\leq \alpha$;
    \item[(ii)] If there exists $F_0\in \F_0$ such that $\mathcal{L}_{F_0}$ is nonempty and there exists $(j,y_0,g_0)$ in $\{1,2\}\times \mathcal{L}_{F_0}$ such that $K_{F_0,jj}(y_0,g_0,y_0,g_0)>0$, then, for any $\alpha \in [0,1/2)$,
    $$\lim_{\eta\rightarrow 0} \limsup_{n\rightarrow\infty} \sup_{F\in \mathcal{F}_0} \E_F[\varphi_{n,\alpha}]= \alpha.$$
	\item[(iii)] If $F\in \mathcal{F}\backslash \mathcal{F}_0$, then $\lim_{n\rightarrow \infty} \E_{F}(\varphi_{n,\alpha})=1$.
\end{enumerate}
\end{theorem}

Theorem \ref{prop:asympt_size} (i) is closely related to Theorem 5.1 and Lemma 2 in AS. It shows that the test $\varphi_{n,\alpha}$ controls the asymptotic size, in the sense that the supremum over $\F_0$ of its level is asymptotically lower or equal to $\alpha$. To prove this result, the key is to establish that, under Assumption~\ref{hyp:regul}, the class of transformed unconditional moment restrictions that characterize the null hypothesis satisfies a manageability condition \citep[see][]{pollard1990empirical}.  Using arguments from \cite{hsu2011consistent}, we then exhibit cases of equality in Theorem \ref{prop:asympt_size} (ii), showing that, under mild additional regularity conditions, the test has asymptotically exact size (when letting $\eta$ tend to zero). Finally, Theorem \ref{prop:asympt_size} (iii), which is based on Theorem 6.1 in AS, shows that the test is consistent over fixed alternatives.

\medskip
\paragraph{Extension to account for aggregate shocks}

This testing procedure can be easily modified to accommodate unanticipated aggregate shocks. Specifically, using the notation defined in Section \ref{ss223}, we consider the same test as above after replacing $\widetilde{Y} $ by $ \widetilde{Y}_{\widehat{c}} = D q(Y,\widehat{c}) + (1-D) \psi$, where $\widehat{c}$ denotes a consistent estimator of $c_0$. The resulting test is given by $\varphi_{n,\alpha, \widehat{c}} = \indic\left\{T(\widehat{c})> c^*_{n,\alpha}\right\}$ (where $T(\widehat{c})$ is obtained by replacing $\widetilde{Y} $ by $\widetilde{Y}_{\widehat{c}}$ in the original test statistic). Such tests have the same properties as those above under some mild regularity conditions on $q(\cdot,\cdot)$, which hold in particular for the leading examples of additive and multiplicative shocks ($q(y,c)=y-c$ and $q(y,c)=y/c$). We refer the reader to Appendix \ref{app:test_shock} for a detailed discussion of this extension.

\section{Monte Carlo simulations}\label{ss32}

In the following we study the finite sample performances of the test without covariates through Monte Carlo simulations. The finite sample performances of the version of our test that accounts for covariates are reported and discussed in Appendix \ref{sec:simus_X}.

\medskip
We suppose that the outcome $Y$ is given by
$$ Y = \rho \psi  + \eps,$$
with $ \rho \in [0,1] $, $\psi \sim \mathcal{N}(0,1)$ and
$$\eps =  \zeta \left(-\indic\{U \leq 0.1\} + \indic\{U \geq 0.9\}\right),$$
where $\zeta$, $U$ and $\psi$ are mutually independent, $\zeta \sim \mathcal{N}(2, 0.1)$ and $U\sim \mathcal{U}[0,1]$. In this setup, $\E(Y|\psi)=\rho\psi$ and expectations are rational if and only if $\rho=1$. But since we observe $Y$ and $\psi$ in two different datasets, there are values of $\rho\neq 1$ for which our test cannot reject the null hypothesis. More precisely, we can show that as the sample size $n$ grows to infinity, we reject the null if and only if $\rho \leq  \rho^* \simeq 0.616$. Besides, given this data generating process, the naive test $\E(Y)=\E(\psi)$ always fails to reject RE, while the RE test based on variances is only able to detect a subset of violations of RE that correspond to $ \rho < 0.445$.

\medskip
To compute our test, we need to choose the tuning parameters $b_0$, $\kappa$, $\epsilon$  and $\eta$ (see Section \ref{sub:stat_tests} for definitions).  As mentioned in Section \ref{sub:stat_tests}, we set $\epsilon=0.05$ and $\eta=10^{-6}$, following \cite{andrews2016inference}.  \cite{andrews2013inference} show that there exists in practice a large range of admissible values for the other tuning parameters parameters. Regarding $b_0$ and $\kappa$, we follow \citeauthor{beare2015improved} (\citeyear{beare2015improved}, Section 4.2) and compute,  for a grid of candidate parameters, the rejection rate under the null and under one alternative (namely, $\rho=0.5$), through Monte Carlo simulations. Then, we set $(b_0,\kappa)$ so as to maximize the power subject to the constraint that the rejection rate under the null is below the nominal size 0.05. That way, we obtain $b_0=0.3$ and $\kappa=0.001$. The parameter $p$ has a distinct effect, in that its choice does not affect size, at least asymptotically. Rather, this parameter selects to what extent the test aims power at the equality constraint $E(Y-\psi)=0$ versus the inequalities $E[(y-Y)^+ - (y-\psi)^+]\geq 0 \ (y\in\R)$. Setting $p$ to $0.05$ leads to slightly higher power in our DGP, but values of $p$ in $[0,0.31]$ provide similar finite sample performances, with power always greater than 90\% of the maximal power.

\begin{figure}[H]
\centering
\includegraphics[width=0.95\linewidth, height=0.3\textheight]{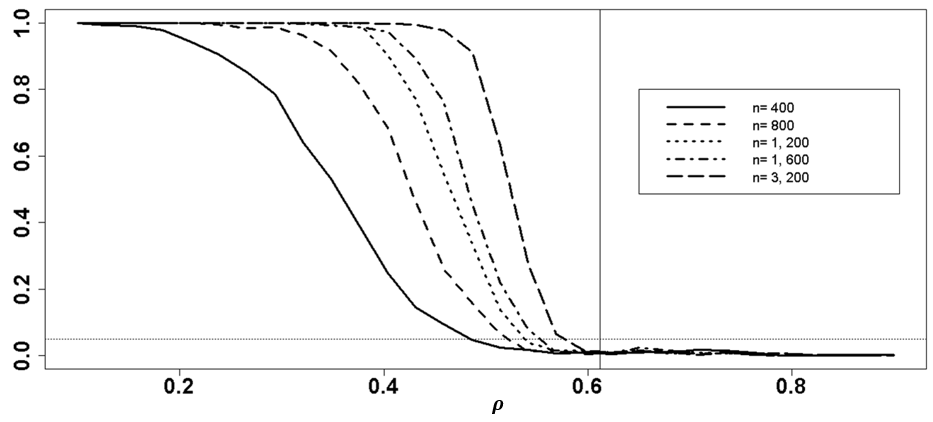}
\begin{minipage}{0.9 \textwidth}
{\footnotesize
Notes: The vertical line at $\rho\simeq 0.616$ corresponds to the theoretical limit for the rejection of the null hypothesis using our test. The dotted horizontal line corresponds to the 5\% level.}
\end{minipage}
\caption{Power curves.}
\label{fig:normal1}
\end{figure}

\medskip
Results reported in Figure \ref{fig:normal1} show the power curves of the test $ \varphi_{\alpha} $ for five different sample sizes ($n_{Y}=n_{\psi}=n \in \{400; 800; 1,200; 1,600; 3,200\}$) as a function of the parameter $\rho$, using 800 simulations for each value of $\rho$. We use 500 bootstrap simulations to compute the critical values of the test. 

\medskip
Several remarks are in order. First, as expected, under the alternative (i.e. for values of $\rho \leq \rho^* = 0.616$), rejection frequencies increase with the sample size $n$. In particular, for the largest sample size $n=3,200$, our test always results in rejection of the RE hypothesis for values of $\rho$ as large as .45. Second, in this setting, our test is conservative in the sense that rejection frequencies under the null are smaller than $\alpha=0.05$, for all sample sizes. This should not necessarily come as a surprise since the test proposed by AS has been shown to be conservative in alternative finite-sample settings (see, \emph{e.g.} Table 1 p.22 in AS for the case of first-order stochastic dominance tests). However, for the version of our test that accounts for covariates and for the data generating process considered in Appendix~\ref{sec:simus_X}, rejection frequencies under the null are very close to the nominal level.

\medskip

\section{Application to earnings expectations}\label{sec4}

\subsection{Data}
\label{subsec:data}

Using the tests developed in Section~\ref{sub:stat_tests}, we now investigate whether household heads form rational expectations on their future earnings.  We use for this purpose data from the Survey of Consumer Expectations (SCE), a monthly household survey that has been conducted by the Federal Reserve Bank of New York since 2012 (see \citeauthor*{armantier2017overview}, \citeyear{armantier2017overview}, for a detailed description of the survey, and \citeauthor{kuchler2015personal}, \citeyear{kuchler2015personal}; \citeauthor*{CPWZ18}, \citeyear{CPWZ18}; \citeauthor*{FKZ18}, \citeyear{FKZ18} for recent articles using the SCE). The SCE is conducted with the primary goal of eliciting consumer expectations about inflation, household finance, labor market, as well as housing market. It is a rotating internet-based panel of about 1,200 household heads, in which respondents participate for up to twelve months.\footnote{Each survey takes on average about fifteen minutes to complete, and respondents are paid \$15 per survey completed.} Each month, the panel consists of about 180 entrants, and 1,100 repeated respondents. While entrants are overall fairly similar to the repeated respondents, they are slightly older and also have slightly lower incomes (see Table 1 in \citeauthor{armantier2017overview}, \citeyear{armantier2017overview}).

\medskip
Of particular interest for this paper is the supplementary module on labor market expectations. This module is repeated every four months since March 2014. Since March 2015, respondents are asked the following question about labor market earnings expectations ($\psi$) over the next four months: ``What do you believe your annual earnings will be in four months?''. Implicit throughout the rest of our analysis is the assumption that these elicited beliefs correspond to the mean of the subjective beliefs distribution.\footnote{This assumption, while often made in the subjective expectations literature, is \textit{a priori} restrictive. In this application, for the vast majority of the sub-groups of the population, the mean of $\psi$ cannot be statistically distinguished from the one of $Y$ (see Table~ \ref{tab:Earnings} below). This provides empirical support for this assumption.} In this module, respondents are also asked about current job outcomes, including their current annual earnings ($Y$), through the following question: ``How much do you make before taxes and other deductions at your [main/current] job, on an annual basis?''.

\medskip
Specifically, we use for our baseline test the elicited earnings expectations ($\psi$), which are available for two cross-sectional samples of household heads who were working either full-time or part-time at the time of the survey, and responded to the labor market module in March 2015 and July 2015 respectively. We combine this data with current earnings ($Y$) declared in July 2015 and November 2015 by the respondents who are working full-time or part-time at the time of the survey.\footnote{Throughout our analysis (with the exception of the number of observations reported in Table~\ref{tab:Earnings}) we use the monthly survey weights of the SCE in order to obtain an estimation sample that is representative of the population of U.S. household heads. See \cite{armantier2017overview} for more details on the construction of these weights. We also Winsorize the top 5 percentile of the distributions of realized earnings and earnings beliefs.} This leaves us with a final sample of 2,993 observations, which is composed of 1,565 earnings expectation observations, and 1,428 realized earnings observations. 51\% (1,536) of these observations correspond to the sub-sample of respondents who are reinterviewed at least once. We refer to Table~\ref{tab:Des} for additional details on our sample.

\begin{table}[H]
       \caption{Descriptive statistics of the SCE sample}
       \centering
       \footnotesize
           \begin{tabular}{@{}lcc@{}}\toprule
     & Mean & Std. dev.     \\ \hline
 Male & 0.53 & 0.50  \\
 White & 0.74 & 0.43 \\
 College degree & 0.49 & 0.46 \\
 Low numeracy & 0.33 & 0.47 \\
 Tenure $\leq$ 6 months  & 0.17 & 0.38  \\
 Age & 45.8 & 13.0    \\
 $\psi$ (Earnings beliefs) &  \$50,592   &   \$40,889     \\
$Y$ (Realized earnings) &   \$52,354     &  \$38,634  \\ \bottomrule \hline
   \multicolumn{3}{p{200pt}}{}
           \end{tabular}\par
           \medskip
           \label{tab:Des}
\end{table}

\subsection{Implementation of the test}

We summarize how we implemented the test in practice, either on the overall sample or on each subsample corresponding to the binary covariates in Table \ref{tab:Des}. For each case, we start by winsorizing the distribution of realized earnings ($Y$) and earnings beliefs ($\psi$) at the 95\% level.\footnote{We show in Table \ref{tab:Winsorization} in Appendix \ref{app:results_RE_test} that our results are robust to other levels of Winsorization.} Then, we perform the test without covariates, where we allow for multiplicative aggregate shock and thus test $H_{0S}$, with $q(y; c) = y/c$.\footnote{In our application, the parameter $c$ is estimated using survey weights from the SCE.} Then, we use the function $\mathtt{test}$ of our companion R package \href{https://CRAN.R-project.org/package=RationalExp}{RationalExp}.\footnote{See Section 3 in our user's guide \citep{RationalExp_R18} for details on this function.} We choose the same values for the tuning parameters $b_0=0.3$ and $\kappa=0.001$ as in the Monte-Carlo simulations in Section \ref{ss32}. We also set $p=0.05$, $\epsilon= 0.05$, and $\eta= 10^{-6}$. Following \cite{andrews2016inference}, the interval $ \widehat{\mathcal{Y}}$ is approximated by a grid of length $100$ from $\underset{i=1,\dots,n}{\min}\widetilde{Y}_i$ to  $\underset{i=1,\dots,n}{\max}\widetilde{Y}_i$. Finally, we use 5,000 bootstrap simulations to compute the critical values of the test.

\subsection{Are earnings expectations rational?}

In Table~\ref{tab:Earnings} below, we report the results from the naive test of RE ($\E(Y)=\E(\psi)$), and our preferred test (``Full RE''), where we allow for multiplicative aggregate shocks. We implement the tests both on the overall population and on separate subgroups. The latter approach allows us not only to identify which groups fail to rationalize RE, but also, and importantly, to account for the possibility that aggregate shocks may in fact differ across subgroups.

\medskip
Several remarks are in order. First, using our test, we reject for the whole population, at any standard level, the hypothesis that agents form rational expectations over their future earnings. Second, we also reject RE (at the 5\% level) when we apply our test separately for whites (non-Hispanics) and minorities, as well as low vs. high numeracy test scores.\footnote{Respondents' numeracy is evaluated in the SCE through five questions involving computation of sales, interests on savings, chance of winning lottery, of getting a disease and being affected by a viral infection. Respondents are then partitioned into two categories: ``High numeracy'' (4 or 5 correct answers), and ``low numeracy'' (3 or fewer correct answers).}

\medskip

Third, the results from our test point to beliefs formation being heterogeneous across schooling (college degree vs. no college degree) and tenure (more or less than 6 months spent in current job) levels. In particular, we cannot rule out that the beliefs about future earnings of individuals with more schooling experience correspond to rational expectations with respect to some information set. Similarly, while we reject RE at any standard level for the subgroup of workers who have accumulated less than 6 months of experience in their current job, we can only marginally reject at the 10\% level RE for those who have been in their current job for a longer period of time. As such, these findings complement some of the recent evidence from the economics of education and labor economics literatures that individuals have more accurate beliefs about their ability as they progress through their schooling and work careers \citep*[see, e.g.,][]{Stinebrickner12,AAMR16}.

\medskip
\begin{table}[H]
	\caption{Tests of RE on annual earnings}
	\footnotesize
	\centering
	\begin{tabular}{@{}lcccccc@{}}\toprule
		& $\E(Y-\psi)/\E(Y)$  &  Naive RE & Variance RE    & Full RE  & \multicolumn{2}{c}{Number of obs.}   \\
		&  &  (p-val) & (p-val)       &   (p-val)   & $\psi$ & $Y$ \\ \midrule
		All &  0.034 &  0.23 &     0.71          & $< 0.001$     &  1,565 & 1,428  \\ \hline
		Women   &  0.059&  0.13 &   0.62       & $< 0.001$       & 730 & 649 \\
		Men   &  0.025    & 0.48 &   0.58     &  0.210    & 835 & 779 \\  \hline
		White &  0.032   &   0.31 &   0.67     & 0.021 & 1,200  & 1,097 \\
		Minorities   & 0.046 &   0.43 &    0.60     & $< 0.006$ & 365&  331 \\  \hline
		College degree &  -0.001 & 0.96&    0.50     & 0.130 & 1,106 & 1,053  \\
		No college degree & 0.093 & 0.04 &    0.57     & 0.013 &459 &  375 \\   \hline
		High numeracy & 0.033 & 0.28&      0.62        & 0.012& 1,158 &1,070  \\
		Low numeracy & 0.055 &  0.27&    0.58     & 0.022 &  407 & 358   \\  \hline
		Tenure $\leq  6$ months & 0.105  &  0.24 &    0.63        & $< 0.001$     & 271 & 180 \\
		Tenure $> 6$ months & 0.007   & 0.81&   0.65     & $0.091$ & 1,294& 1,248\\
		\bottomrule
		\multicolumn{7}{p{390pt}}{{\footnotesize Notes: ``Naive RE'' denotes the naive RE test of equality of means between $Y$ and $\psi$. ``Variance RE'' denotes the variance RE test where the null hypothesis is the variance of $Y$ being greater or equal than the variance of $\psi$, once we account for aggregate, multiplicative shocks. ``Full RE" denotes the test without covariates, where we test $\text{H}_{0S}$ with $q(y,c)=y/c$. We use 5,000 bootstrap simulations to compute the critical values of the Full RE test. Distributions of realized earnings ($Y$) and earnings beliefs ($\psi$) are both Winsorized at the 95\% quantile.}} 
	\end{tabular}
	\label{tab:Earnings}
\end{table}

Fourth, using the naive test of equality of means between earnings beliefs and realizations, one would instead generally not reject the null at any standard levels. The one exception is the subgroup of workers without a college degree, for whom the naive test yields rejection of RE at the 5\% level. But, as discussed before, one cannot rule out that such a rejection is due to aggregate shocks.

\medskip
Even though individuals in the overall sample form expectations over their earnings in the near future that are realistic, in the sense of not being significantly biased, the result from our preferred test shows that earnings expectations are nonetheless not rational. Taken together, these findings highlight the importance of incorporating the additional restrictions of rational expectations that are embedded in our test, using the distributions of subjective beliefs and realized outcomes to detect violations of rational expectations. That the variance test of RE never rejects the null at any standard levels indicates that it is important in practice to go beyond the first moments, and exploit instead the full distributions of beliefs and outcomes to detect departures from rational expectations. These results also suggest that, in order to rationalize the realized and expected earnings data, one should consider alternative models of expectation formation that primarily differ from RE in their third, or higher-order moments.

\medskip
The results of the direct test of RE on the subsample of individuals who are followed over four months are reported in Table \ref{tab:direct} below. While these results generally paint a similar picture to the results of our test, there are some differences. In particular, the direct test rejects RE at the 5\% level for men and at 1\% for individuals with tenure greater than 6 months, whereas we do not reject RE for the former group and only marginally so, at the 10\% level, for the latter. The direct test also rejects with less power than our test for certain groups (low numeracy, tenure lower than 6 months, and minorities). This lower power may seem surprising given that the direct test can exploit the joint distribution of $(Y,\psi)$, but is simply due to the important reduction in sample size when focusing on the subsample of individuals who are followed over four months results.

\medskip
There are also important issues associated with the direct test, which generally warrant caution when interpreting the results from this test. Most importantly, as already discussed in Section~\ref{sub:meas_err}, the direct test is not robust to measurement errors on the subjective beliefs $\psi$. As shown in Proposition  \ref{prop:reg_meas_err} in Appendix \ref{sec:lin_reg_meas_err}, it is however possible to derive a restriction on $\beta$ under RE. Specifically, if $\xi_\psi$ is positively correlated with $\eps + \xi_Y$, we have, under RE,
\begin{equation}
\beta\geq 1 - \frac{1}{1+\underline{\lambda}},	
	\label{eq:lower_bound_beta}
\end{equation}
where $\underline{\lambda}$ is a lower bound on the signal-to-noise ratio $\V(\psi)/\V(\xi_\psi)$. Table \ref{tab:direct} also reports the results of tests combining \eqref{eq:lower_bound_beta}  with the restrictions on the marginal distributions used in our full RE test. Adding the restriction \eqref{eq:lower_bound_beta}  does not change the results for values of signal-to-noise ratio between 5 and 20 (i.e., for noise-to-signal ratios between 5\% and 20\%). Overall, using the subsample of linked data $(Y,\psi)$ through this additional restriction does not add much to our test, at least once we account for possible measurement errors on the elicited beliefs. Another significant concern with the direct test, and, more generally, the use of linked data on $(Y,\psi)$, is that attrition may be endogenous. We discuss this issue in more details in Appendix \ref{app:results_RE_attrition}.

\medskip
 \begin{table}[H]
	\caption{Direct test, our test, and combined test of RE on annual earnings}
	\scalebox{0.90}{
	\footnotesize
	\centering
	\begin{tabular}{@{}lcccccccc@{}}\toprule
		& $\beta$  &  Direct test   & Full RE  & \multicolumn{2}{c}{Combined test}    & \multicolumn{3}{c}{Number of obs.}   \\
			\cmidrule{5-6}
		Bound on signal/noise $\underline{\lambda}$&  &  & & 5 & 20 & & &   \\
		Implied bound on $\beta$ &  & & &  0.833   & 0.952  & & &        \\
		&  &  (p-val) & (p-val) &  (p-value)  &   (p-value)    & $\psi$ & $Y$ & $(\psi,Y)$ \\ \midrule
		All & 0.954 &  $0.001$  & $< 0.001$ & $< 0.001$  &  $< 0.001$     &  1,565 & 1,428  & 768  \\ \hline
		Women   &  0.956  &  $0.002$  &  $< 0.001$  & $< 0.001$  &      $< 0.001$   & 730 & 649& 356  \\
		Men   &  0.960    & $0.021$ &  0.210 &0.276   &0.276     & 835 & 779 & 412 \\  \hline
		White       &  0.963   &   $0.004$  &  0.021 & 0.019   &   $0.010$   & 1,200  & 1,097& 596 \\
		Minorities   & 0.928 &   $0.010$ & 0.006& 0.007     & 0.005   & 365&  331& 172 \\  \hline
		College degree &  0.974 & $0.060$ & 0.130&   0.182    &  0.182    & 1,106 & 1,053 & 560 \\
		No college degree & 0.954 &   $0.044$ & 0.013 &  0.017    & 0.017    &459 &  375& 208  \\   \hline
		High numeracy & 0.959 & $0.001$  & 0.012&  0.016    &  0.016   & 1,158 &1,070 & 573 \\
		Low numeracy & 0.954 & $0.094$  & 0.022 & 0.030    &   0.030    &  407 & 358 &  195 \\  \hline
		Tenure $\leq  6$ months & 0.942  &  $0.015$ & $0.001$ & 0.002  &  0.001       & 271 & 180& 98 \\
		Tenure $> 6$ months & 0.956   & $0.001$ & $0.091$&  0.094    &  0.094  & 1,294& 1,248 & 670\\
		\bottomrule
		\multicolumn{9}{p{450pt}}{{\footnotesize Notes: ``Direct test'' denotes the direct test of RE when $(\psi,Y)$ is observed. $\beta$ is the coefficient of the regression of $Y$ on $\psi$ in that case.  ``Full RE" denotes the test without covariates, where we test $\text{H}_{0S}$ with $q(y,c)=y/c$. We use 5,000 bootstrap simulations to compute the critical values of the Full RE test. ``Combined RE test'' denotes the test without covariates, where we test $\text{H}_{0S}$ with $q(y,c)=y/c$, which is the ``Full RE" test,  combined with the additional restriction $\beta \geq  1-1/(1+\underline{\lambda})$, where $\underline{\lambda}$ is an a priori bound on the signal-to-noise ratio.  Distributions of realized earnings ($Y$) and earnings beliefs ($\psi$) are both Winsorized at the 95\% quantile.}} %Income 1: CMS($\psi$)/ CMS(Y) and 1990-2016, Income 2:  CMS($\psi$)/ SIPP(Y) and 2005-2009}
	\end{tabular}
}
	\label{tab:direct}
\end{table}

\medskip
Coming back to our test, the rejection of RE for the overall population but also for most of the subpopulations are, in view of Proposition \ref{prop:meas_err}, unlikely to be due to data quality issues. In that sense, these results may be seen as robust evidence against the RE hypothesis for individual earnings, at least in this context. As a result, conclusions of behavioral models based on the assumption that agents form rational expectations about their future earnings may be misleading. Exploring this important question requires one to go beyond testing though, by quantifying the extent to which model predictions are actually sensitive to the violations from rational expectations that have been detected with our test. We investigate this issue in \cite{DGM} in the context of a life-cycle consumption model.

\section{Conclusion}\label{sec:concl}

In this paper, we develop a new test of rational expectations that can be used in a broad range of empirical settings. In particular, our test only requires having access to the marginal distributions of realizations and subjective beliefs. As such, it can be applied in frequent cases where realizations and beliefs are observed in two separate datasets, or only observed for a selected sub-population. By bypassing the need to link beliefs to future realizations, our approach also enables to test for rational expectations without having to wait until the outcomes of interest are realized and made available to researchers. We establish that whether one can rationalize rational expectations is equivalent to the distribution of realizations being a mean-preserving spread of the distribution of beliefs, a condition which can be tested using recent tools from the moment inequalities literature. We show that our test can easily accommodate covariates and aggregate shocks, and, importantly for practical purpose, is robust to some degree of measurement errors on the elicited beliefs. We apply our method to test for rational expectations about future earnings, using data from the Survey of Consumer Expectations. While individuals tend to be right on average about their future earnings, our test strongly rejects rational expectations. 

\medskip
Beyond testing, in this application as in any other situations where rational expectations are violated, a natural next step is to evaluate the deviations from rational expectations that one can rationalize from the available data. In the context of structural analysis, a central question then becomes to which extent the main predictions of the model are sensitive to those departures from rational expectations. We explore this important issue and propose in \cite{DGM} a tractable sensitivity analysis framework on the assumed form of expectations.

\newpage
\bibliography{../../bib_Roy5}

\newpage
\appendix

\section{Aggregate shocks}

\subsection{Statistical tests in the presence of aggregate shocks}	\label{app:test_shock}

In this appendix, we show how to adapt the construction of the test statistic and obtain similar results as in Theorem \ref{prop:asympt_size} in the presence of aggregate shocks. As explained in Section \ref{ss223}, we mostly have to replace $\widetilde{Y}$ by $\widetilde{Y}_{c}= Dq\left(\widetilde{Y}, c \right) + (1-D)\psi$. Because we include covariates here, as in Section \ref{sub:stat_tests}, $c$ is actually a function of $X$. Also, the true function $c_0$ has to be estimated. We let $\widehat{c}$ denote such a nonparametric estimator, which is based on $\E[q(Y,c_0(X))|X]=\E[\psi|X]$. When $q(y,c)=y-c$ or $q(y,c)=y/c$, we get respectively $c_0(X) = \E(Y|X)-\E(\psi|X)$ and $c_0(X) = \E(Y|X)/\E(\psi|X)$, and $\widehat{c}$ is easy to compute using nonparametric estimators of $\E(Y|X)$ and $\E(\psi|X)$.

\medskip
Because in Proposition \ref{prop:equiv_gen} (ii) we do not test for a moment equality anymore, $m\left(D_i,\widetilde{Y}_i, X_i, g,y\right)$ reduces to $ m_1\left(D_i,\widetilde{Y}_{c,i}, X_i, g,y\right)$. We let hereafter $\overline{m}_n(g,y)= \sum_{i=1}^n m_1\left(D_i,\widetilde{Y}_{c,i}, X_i, g,y\right) /n$. In the test statistic $T$, we replace, for $ (y,g) \in \mathcal{Y}\times \cup_{r\geq 1}\mathcal{G}_r $, $\overline{\Sigma}_n(g,y)$ by $ \overline{\Sigma}_n(g,y) = \widehat{\Sigma}_n(g,y) + \epsilon \mathrm{Diag}\left( \widehat{\mathbb{V}}\left(\widetilde{Y}_{\hat{c}} \right)    ,  \widehat{\mathbb{V}}\left(\widetilde{Y}_{\hat{c}}  \right)     \right)$,  where  $\widehat{\Sigma}_n(g,y)$ and $\widehat{\mathbb{V}}\left(\widetilde{Y}_{\hat{c}} \right) $ are respectively the sample covariance matrix of $ \sqrt{n}\overline{m}_n\left( g,y\right)$ and the empirical variance of $\widetilde{Y}_{\hat{c}}$. The last difference with the test considered in Section \ref{sub:stat_tests} is that when using the bootstrap to compute the critical value, we also have to re-estimate $c_0$ in the bootstrap sample.

\medskip
We obtain in this context a result similar to Theorem \ref{prop:asympt_size} above, under the regularity conditions stated in Assumption \ref{hyp:gen}. We let hereafter $ \mathcal{C}_s\left([0,1]^{d_X}\right) $ denote the space of continuously differentiable functions of order $ s $ on $[0,1]^{d_X}$ that have a finite norm
$ \left\| c\right\|_{s,\infty} = \underset{\abs{\mt{k}}\leq s}{\max} \sup_{x\in [0,1]^{d_X}}\left|c^{(\mt{k})}(x)\right|$. We also let, for any function $f$ on a set $ \mathcal{G} $, $ \left\| f \right\|_{\mathcal{G}} = \sup_{x\in\mathcal{G}}\left| f(x) \right|  $.  Finally, when the distribution of $\left(D,\widetilde{Y},X\right)$ is $F$, $K_F$ denotes the asymptotic  covariance kernel of $n^{-1/2}\mathrm{Diag}\left(\mathbb{V}\left(\widetilde{Y}_{c_0}\right)\right)^{-1/2} \overline{m}$.

\begin{assumption}\label{hyp:gen}
\begin{enumerate}
\item[(i)] $\widehat{c}$ and $ c_0 $ belong to $ \mathcal{C}_s\left([0,1]^{d_X}\right)$, with $ s \geq d_X$.
 Moreover, $\|\widehat{c}-c_0\|_{[0,1]^{d_X}}=o_P(1)$.
\item[(ii)] For all $ y \in \mathcal{Y} $, $q$ is Lipschitz on $ \mathcal{Y}\times[-C,C] $ for some $C>\|c_0\|_{[0,1]^{d_X}}$. Moreover, $ \sup_{(y,c)\in\mathcal{Y}\times[-C,C]} \abs{q(y,c)} \leq M_0$;
\item[(iii)] For all $ c \in \R $, the function $ q(\cdot, c): \mathcal{Y} \to \mathcal{Y}  $ is bijective and its inverse $ q^{I}(\cdot, c) $ is Lipschitz on $ \mathcal{Y} $;
\item[(iv)] $ F_{\psi|X}(\cdot|x)$,  $F_{Y|X}(\cdot|x)$ are Lipschitz on $ \mathcal{Y} $ uniformly in $x\in [0,1]^{d_X}$  with constants $Q_{F,1}$ satisfying $\sup_{F\in \mathcal{F}_0}Q_{F,1} \leq  \overline{Q}_{1}< \infty$. Also, $ F_{q\left(\psi,c(X)\right)}$, $F_{q\left(Y,c(X)\right)} $ are  Lipschitz on $ [-M_0, M_0] $ with constants $Q_{F,2}$ satisfying $\sup_{F\in \mathcal{F}_0}Q_{F,2} \leq  \overline{Q}_{2}< \infty$;
\item[(iv)] $\inf_{F\in \mathcal{F}}\mathbb{V}_F\left[\widetilde{Y}_{c}^{2}\right]>0$ and $\epsilon_0 \leq\inf_{F\in \mathcal{F}} \E_F\left[ D \right] \leq \sup_{F\in \mathcal{F}}  \E_F\left[ D \right]\leq 1-\epsilon_0$ for some $\eps_0\in (0,1/2)$. Also,  $\widehat{\mathbb{V}}_F\left[\widetilde{Y}_{\widehat{c}}^{2}\right] $ is a consistent estimator of $\mathbb{V}_F\left[\widetilde{Y}_{c}^{2}\right]$.
\end{enumerate}
\end{assumption}
Part (i) imposes some regularity conditions on $c_0$ and its nonparametric estimator $\widehat{c}$. It is possible to check such regularity conditions on $\widehat{c}$ with kernel or series estimators of $\E(Y|X)$ and $\E(\psi|X)$. Parts (ii) and (iii) also hold when $q(y,c)=y-c$ and $q(y,c)=q(y)/c$, by imposing in the second case that $c$ belongs to a compact subset of $(0,\infty)$. Proposition \ref{prop:asympt_size_gen} shows that under these conditions, the test has  asymptotically correct size.

\begin{proposition}\label{prop:asympt_size_gen}
Suppose that $r_n\rightarrow \infty$ and that Assumptions \ref{hyp:regul} and \ref{hyp:gen} hold. Then (i) in Proposition \ref{prop:asympt_size} holds, replacing $ \varphi_{n,\alpha} $ by $ \varphi_{n,\alpha, \widehat{c}} $.
\end{proposition}

Results like (ii) and (iii) in Proposition \ref{prop:asympt_size} could also be obtained under the conditions of Proposition \ref{prop:asympt_size_gen}, modifying directly the proof of Proposition \ref{prop:asympt_size}.

\subsection{Impossibility results with more flexible effects of aggregate shocks}	
\label{subsec:imp_shock}

We show here that restrictions in the way aggregate shocks affect the outcome are needed to be able to reject RE with $F_Y$ and $F_\psi$. We consider for that purpose the following model:
\begin{equation}
Y= \sum_{k=0}^K C_k V^k + \eps,
\label{eq:AR1}
\end{equation}
where $V$  is $\mathcal{I}$-measurable and the individual shock $\eps$ satisfies $E[\eps|\mathcal{I}]=0$. The vector $C:=(C_0,...,C_K)'$ represents aggregate shocks, which is assumed to be independent of $\mathcal{I}$, with support $\R^{K+1}$. We also assume that $E(C)=(0,1,0,...,0)'$, so that $V=\E[Y|\mathcal{I}]$ and under RE, $\psi=V$. Let $Q_c(y)= \sum_{k=0}^K c_k y^k$. Then $\E(Y|C=c,\mathcal{I})=Q_c(V)$ and under RE, we have
$$\E(Y|C=c,\mathcal{I})=Q_c(\psi).$$
Hence, as in Section \ref{ss223}, we consider the following hypothesis:
\begin{align*}
\text{H}_{0SK}:& \; \text{there exist random variables }  \left(Y', \psi'\right), \text{ a sigma-algebra } \mathcal{I}' \text{ and } c\in\R^{K+1} \text{ such that } \\
 & \sigma(\psi') \subset \mathcal{I}', \, Y' \sim Y, \, \psi'\sim \psi \text{ and } \E\left[Y'\middle|\mathcal{I}'\right]=Q_c(\psi').
\end{align*}

The following proposition is a negative result on the possibility to test for $\text{H}_{0SK}$. 

\begin{proposition}
Suppose that $F_Y$ and $F_\psi$ are continuous with supports that are bounded intervals. For any $\eta>0$,  there exists $K>0$ and $F$, with $\sup_{u\in\R} |F(u)-F_{\psi}(u)|<\eta$, such that $H_{0SK}$ holds with $Y$ and $\widetilde{\psi}\sim F$ (instead of $\psi$).
\label{prop:imposs_result} 
\end{proposition}

Proposition \ref{prop:imposs_result} states  that as $K$ grows large, the set of cdfs $F_Y$ and $F_\psi$ satisfying $\text{H}_{0SK}$ (and thus RE in Model \eqref{eq:AR1}) becomes arbitrarily close, for the Kolmogorov-Smirnov metric, to the set of  of cdfs $F_Y$ and $F_\psi$ that do not satisfy $H_{0SK}$. In other words, $\cup_{K\in\N} \text{H}_{0SK}$ is dense in the set of all continuous cdfs having bounded interval as supports. When combined with Theorem 2 in \cite{bertanha2020impossible}, this implies that there does not exist any almost-surely continuous test of $\cup_{K\in\N} \text{H}_{0SK}$ that has non-trivial power. 

\medskip
A similar, negative result holds if aggregate shocks are allowed to vary with respect to unobserved, individual-specific variables. For instance, shocks may be sector-specific, but sectors may be unobserved in the data. To show such an impossibility result, consider the following model:  
$$Y = q(C,U) + V + \eps,$$
where both $U$ and $V$ are $\mathcal{I}-$measurable, $C$ is an aggregate shock independent of $\mathcal{I}$ and the individual shock $\eps$ satisfies $E[\eps|\mathcal{I}]=0$. Thus, aggregate shocks affect the outcome in an additive way, but heterogeneously across individuals, depending on their $U$, which is assumed to be unobserved by the econometrician and can thus depend on $V$ in a flexible way. We assume without loss of generality that $E[q(C,U)|\mathcal{I}]=0$, so that $\psi=V$ under RE. Let us also assume that $q(u,c)=\sum_{k=0}^K c_k u^k$ and $U=\xi V$, with $\xi>0$, $\xi\indep V$ and $\E[\xi^k]<\infty$ for all $k\leq K$. Let $C'_k=E[\xi^k] C_k$ if $k\neq 1$, $C'_1 = E[\xi] C_1 - 1$ and $C'=(C'_0,...,C'_K)'$. Then, under RE,
$$\E[Y|C'=c',\mathcal{I}] = \sum_{k=0}^K c'_k \psi^k.$$
Moreover, if $\Supp(C)=\R^{K+1}$, we also have $\Supp(C')=\R^{K+1}$, and no constraint is imposed on $c'$.\footnote{$E[q(C,U)|\mathcal{I}]=0$ implies that $E[C_k]=0$ for $k=0,...,K$, but it does not restrict the set of possible $c'_k$.} As a result, we are led again to test $H_{0SK}$, and the same negative result as above holds.

\section{Tests based on linear regressions with measurement errors}
\label{sec:lin_reg_meas_err}

We suppose here to observe both $(\widehat{Y},\widehat{\psi})$ satisfying \eqref{eq:meas_errors}. In this framework, we study the restrictions that RE entail on the coefficient $\beta$ of the (theoretical) linear regression of $\widehat{Y}$ on $\widehat{\psi}$.

\begin{proposition}\label{prop:reg_meas_err}
	\begin{enumerate}
		\item For any values of $(\V(\widehat{Y}), \V(\psh),\Cov(\Yh,\psh))$ such that $\V(\Yh)>\V(\psh)$, there exists a DGP compatible with this triple, satisfying \eqref{eq:meas_errors}, for which RE hold and such that $\eps+\xi_Y \indep \psi$ and $F_{\xi_\psi}$ dominates at the second order $F_{\xi_Y+\eps}$.
		\item If $\beta < 1 -1/(1+\underline{\lambda})$ for some $\underline{\lambda}\geq 0$,  there exists no DGP compatible with this value of $\beta$, satisfying \eqref{eq:meas_errors}, for which RE hold and such that corr$(\xi_\psi,\xi_Y+\eps)\geq 0$ and $\V(\psi)/\V(\xi_\psi)\geq \underline{\lambda}$.
	\end{enumerate}
\end{proposition}

The first result is a negative one. It implies that without further restrictions than those already imposed in Proposition \ref{prop:meas_err}, the regression of $\widehat{Y}$ on $\widehat{\psi}$ does not bring any additional restriction related to RE. The second result, on the other hand, shows that if one assumes a positive correlation between $\xi_\psi$ and $\xi_Y+\eps$ and a lower bound on the signal-to-noise ratio $\V(\psi)/\V(\xi_\psi)$, then $\beta$ is bounded from below under RE. The restriction corr$(\xi_\psi,\xi_Y+\eps)\geq 0$ seems reasonable. First, given that the shocks $\eps$ cannot be anticipated, it is natural to assume that corr$(\xi_\psi,\eps)=0$. It then follows that the assumption corr$(\xi_\psi,\xi_Y+\eps)\geq 0$ holds if the measurement errors on $Y$ and $\psi$ are positively correlated. This would typically happen, for instance, if individuals report their expectations and realized earnings omitting in both cases some components of their earnings, or if they instead overstate their realized earnings, and their expectations accordingly. 

\medskip
This proposition just focuses on the linear regression of $\widehat{Y}$ on $\widehat{\psi}$, since this regression has been very often used to test for RE. This means, however, that there may in principle be additional restrictions on the joint distribution of $(\widehat{Y},\widehat{\psi})$ implied by RE.

\section{Tests with rounding practices}
\label{sec:test_bounds}

We have considered in Section \ref{sub:meas_err} the possibility of measurement errors on $\psi$. Another source of uncertainty on $\psi$ is rounding. Rounding practices by interviewees are common. A way to interpret these practices is that in situations of ambiguity, individuals may only be able to bound the distribution of their future outcome $Y$ \citep[][]{Manski2004}. If individuals round at 5\% levels, for instance, an answer $\psi=0.05$ for the beliefs about percent increase of income should then only be interpreted as $\psi \in [0.025,0.075]$. Another case where only bounds on $\psi$ are observed is when questions to elicit subjective expectations take the following form: ``What do you think is the percent chance that your own [$Y$] will be below [$y$]?'', for a certain grid of $y$. If 0\ and 100\ are always observed, or if we assume that the support of subjective distributions is included in $[\underline{y},\overline{y}]$, we can still compute bounds on $\psi$.\footnote{Note however that in this case, our approach does not take into account all the information on the subjective distribution.} In such cases, we only observe $(\psi_L ,\psi_U)$, with $\psi_L  \leq \psi \leq  \psi_U$. For a thorough discussion of this issue, and especially of how to infer rounding practices, see \cite{manski2010rounding}.

\medskip
In this setting, rationalizing rational expectations is less stringent than in our baseline set-up since the constraints on the distribution of $\psi$ are weaker. Formally, the null hypothesis takes the following form:
$$\text{H}_{0B}:  \; \exists (Y',\psi', \mathcal{I}'): \; \sigma(\psi')\subset \mathcal{I'}, \; Y'\sim Y, \; F_{\psi_U} \leq F_{\psi'} \leq F_{\psi_L} \text{ and } \E(Y'|\mathcal{I}')=\psi'.$$
To obtain an equivalent formulation to H$_{0B}$, a natural idea would be to fix a candidate cdf $F\in[ F_{\psi_U}, F_{\psi_L}]$ for $F_\psi$ and apply Theorem \ref{prop:equiv} with this $F$. Then, letting $\Delta_F(y)=\int_{-\infty}^y   F_{Y} (t) - F(t) dt$ and $\delta_F=\E(Y)-\int udF(u)$, H$_{0B}$ would hold as long as for some $F\in[ F_{\psi_U}, F_{\psi_L}]$, $\Delta_F(y)\geq 0$ for all $y\in\R$ and $\delta_F=0$. In practice though, directly checking whether such a distribution exists would be very difficult. Fortunately, we show in the following proposition that it is in fact sufficient to check that these conditions hold for a specific candidate distribution. To define the cdf of this distribution, we introduce, for all $b\in \R$, the random variables $$\psi^{b} = \psi_{U} \indic\{\psi_U< b\} + \max(b,\psi_L) \indic\{\psi_U\geq b\}.$$
We also let $\psi^{-\infty}=\psi_L$ and $\psi^{\infty}=\psi_U$. The cdf of $\psi^b$ is then $ F^b(t) =  F_{\psi_U}(t) \indic\{t< b\} +   F_{\psi_L}(t) \indic\{t\geq b\}$, for all $b\in \overline{\R}$. We let $\mathcal{F}_B=\{F^b, b\in \overline{\R}\}$ denote the set of all such cdfs.

\begin{assumption}\label{hyp:regul_equiv_bounds}
	$\E(|Y|)<\infty$, $\E(|\psi_L|)<\infty$ and $\E(|\psi_U|)<\infty$.
 \end{assumption}

\begin{proposition}\label{prop:equiv_bounds}
 Suppose that Assumption \ref{hyp:regul_equiv_bounds} holds. First, if $\E[\psi_L]\leq \E[Y]\leq \E[\psi_U]$, there exists a unique $F^*\in \mathcal{F}_B$ such that $\delta_{F^*}=0$. Second, the following statements are equivalent:
\begin{enumerate}
\item[(i)] $\text{H}_{0B}$ holds.
\item[(ii)] $\E[\psi_L]\leq \E[Y]\leq \E[\psi_U]$ and $\Delta_{F^*} (y) \geq  0 $ for all $y\in \R$.
\end{enumerate}
\end{proposition}

This test shares some similarities with the test in the presence of aggregate shocks. Specifically, if $\E[\psi_L]\leq \E[Y]\leq \E[\psi_U]$, we first identify $b_0\in \overline{\R}$ such that the candidate belief $\psi^{b_0}$, which plays a similar role as the modified outcome $q(Y,c_0)$ in the test with aggregate shocks, satisfies the equality constraint $\E[\psi^{b_0}]=\E[Y]$. Noting that the inequality $\Delta_{F^*} (y) \geq  0 $ can be rewritten as $\E\left[    \left(y-Y\right)^+ -  \left(y-\psi^{b_0}\right)^+\right] \geq 0 $, it follows from (ii) that rationalizing RE in this context (i.e., $\text{H}_{0B}$) is then equivalent to a set of many moment inequality constraints involving the distributions of realizations $Y$ and candidate belief $\psi^{b_0}$.

\section{Tests with sample selection in the datasets}
\label{sec:propensity_score}

We consider here cases where the two samples are not representative of the same population, or formally, $D$  is not independent of $(Y,\psi)$. This may arise for instance because of oversampling of some subpopulations or differences in nonresponse between the two surveys that are used. We assume instead that selection is conditionally exogenous, that is to say:
\begin{equation}
D\indep (Y,\psi)|X.
	\label{eq:CIA}
\end{equation}
We show how to use a propensity score weighting to handle such a selection. Denote by $ p(x) = P(D=1|X=x) = \E\left[D \middle|X=x\right] $ the propensity score and by
\[ W(X) = \df{D}{p(X)}- \df{1-D}{1-p(X)} . \]
The law of iterated expectations combined with Proposition \ref{prop:equivX} directly yields the following proposition:

\begin{proposition}\label{prop:test_sel}
Suppose that \eqref{eq:CIA} and Assumption \ref{hyp:regul_equiv} hold. Then $\text{H}_{0X}$ is equivalent to
$$  \E\left[ W(X) \left(y-\widetilde{Y}\right)^+ \middle| X\right] \geq 0  $$ for all $y\in \R$  and  $ \E\left[ W(X) \widetilde{Y}\middle|X\right]=0$.
\end{proposition}

This proposition shows that under sample selection, we can build a statistical test of $\text{H}_{0X}$ akin to that developed in Section \ref{sub:stat_tests}, by merely estimating nonparametrically $p(X)$. We could consider for that purpose a series logit estimator, for instance. Validity of such a test would follow using very similar arguments as for the test with aggregate shocks considered above.

\section{Simulations with covariates}
\label{sec:simus_X}

We consider here simulations including covariates. The DGP is similar to that considered in Section \ref{ss32}. Specifically, we assume that $ Y = \rho \psi  + \sqrt{X}\eps$, with $ \rho \in [0,1] $, $\psi \sim \mathcal{N}(0,1)$, $X\sim \text{Beta}(0.1, 10)$ and
$$\eps =  \zeta \left(-\indic\{U \leq 0.1\} + \indic\{U \geq 0.9\}\right),$$
where $\zeta \sim \mathcal{N}(2, 0.1)$ and $U\sim \mathcal{U}[0,1]$. $(\psi, \zeta, U, X)$ are supposed to be mutually independent. Like in the test without covariates, we can show that the test with covariates is able to reject RE if and only if $\rho< 0.616$. On the other hand, $\E\left[Y|X\right]=\E\left[\psi|X\right]$, so the naive conditional test has no power. The test based on conditional variances rejects only if  $\rho <0.445$. Finally, we can show that without using $X$, our test has power only for $\rho < 0.52$. Hence, relying on covariates allows us to gain power for $\rho \in [0.521, 0.616)$.

\medskip
Again, we consider $n_{\psi}=n_Y=n\in \{400; 800;  1,200; 1,600; 3,200\}$, use 500 bootstrap simulations to compute the critical value, and rely on 800 Monte-Carlo replications for each value of $\rho$ and $n$. We use the same parameters $ p=0.05 $ and $b_0=0.3$ as above.

\medskip
\begin{figure}[H]
\centering
\includegraphics[width=0.95\linewidth, height=0.30\textheight]{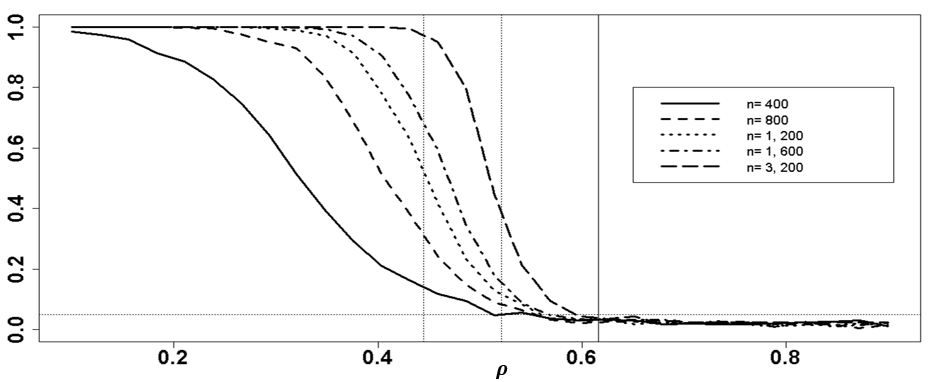}
\begin{minipage}{0.9 \textwidth}
{\footnotesize
Notes: the dotted vertical lines correspond to the theoretical limit for the rejection of the null hypothesis for test based on variance $(\rho\simeq 0.445)$, our test without covariates ($\rho \simeq 0.521$) and our tests with covariates ($\rho=0.616$). The dotted horizontal line corresponds to the 5\% level.}
\end{minipage}
\caption{Power curves for the test with covariates.}
\label{fig:MCratXs}
\end{figure}

Figure \ref{fig:MCratXs} shows that the RE test with covariates asymptotically outperforms the RE test without covariates. The test exhibits a similar behavior as that without covariates, though, as we could expect, the power converges less quickly to one as $n$ tends to infinity.

\section{Additional material on the application}
\label{app:dec}

\subsection{Effect of the Winsorization on the RE test}
\label{app:results_RE_test}

\vspace{-0.7cm}

 \begin{table}[H]
	\caption{Full test of RE with different levels of Winsorization}
	\footnotesize
	\centering
	\begin{tabular}{@{}lccc@{}}\toprule
		Winsorization level &  0.95& 0.97& 0.99   \\
		& (p-value)  &  (p-value) & (p-value)  \\ \midrule
		All &$<$ 0.001 &    $<$ 0.001  &     0.002  \\ \hline
		Women   &   $<$ 0.001  & $<$ 0.001 &   0.001  \\
		Men   & 0.210     & 0.254 &  0.342    \\  \hline
		White &   0.021 &    0.030&  0.049  \\
		Minorities   & 0.006 &  0.007 &     0.018\\  \hline
		College degree &  0.130 & 0.146&     0.196  \\
		No college degree & 0.013& 0.012 &   0.009 \\   \hline
		High numeracy & 0.012  & 0.017&    0.034 \\
		Low numeracy & 0.022 & 0.026&    0.029 \\  \hline
		Tenure $\leq  6$ months &0.001 &  0.005&   0.009   \\
		Tenure $> 6$ months &  0.091   &  0.118&    0.304\\
		\bottomrule
		\multicolumn{4}{p{300pt}}{{\footnotesize Notes: We test $\text{H}_{0S}$ with $q(y,c)=y/c$, using 5,000 bootstrap simulations to compute the critical values. Distributions of realized earnings ($Y$) and earnings beliefs ($\psi$) are both Winsorized at either the 0.95, 0.97, or 0.99 quantile.}} %Income 1: CMS($\psi$)/ CMS(Y) and 1990-2016, Income 2:  CMS($\psi$)/ SIPP(Y) and 2005-2009}
	\end{tabular}
	\label{tab:Winsorization}
\end{table}

\subsection{Possibly endogenous attrition in the survey}
\label{app:results_RE_attrition}

In addition to measurement errors, another potential issue when using the linked data $(Y,\psi)$ is that attrition may be related to $Y$ itself. This would create a sample selection issue that would invalidate the direct test, even absent any measurement errors. To explore this possibility, Table \ref{tab:Earn} below reports the estimation results from a logit model of attrition on earnings beliefs, gender, race/ethnicity, college degree attainment, numeracy test score, tenure and a (linear) time trend.  The main takeaway from this table is that earnings beliefs $\psi$ are significantly associated with attrition, even after controlling for this extensive set of characteristics. This result suggests that individuals for whom we observe both earnings expectations and realizations are likely to earn more than those who are not followed across the two waves. Along the same lines, a Kolmogorov-Smirnov test rejects at the 1\% level the equality of the distributions of realized earnings between the whole sample and the subsample that would be used for the direct test. Similarly, we reject the equality of the distributions of expected earnings between these two samples. These results indicate that, in this context, the direct RE test is likely to be misleading. Conversely, attrition is unlikely to be an issue with our test, since we use in each wave the observations of all respondents.\footnote{The one assumption we need to make is that respondents in the surveys used to measure $\psi$ (i.e., those of March and July 2015) are drawn from the same population as those from the surveys used to measure $Y$ (i.e., those of July and November 2015). That there is no significant time trend in the attrition model (Table~\ref{tab:Earn}) suggests that this assumption is reasonable in this context.}

\begin{table}[H]
	\caption{Logit model of attrition}
	\centering
	\footnotesize
	\begin{tabular}{@{}lcccccccccl@{}}\toprule
	   & Intercept & $\psi$   & Male &  White & Coll. Degree  & Low Num. &    Tenure $> 6$  &  Trend \\  \midrule
		All     &  1.327${}^{\ast\ast}$ &    -6.206e-06${}^{\ast\ast}$&   0.046 & -0.311 &  -0.137 &   -0.141 &   -0.786${}^{\ast\ast}$ &   -0.040	\\
		&  (0.293) & (1.621e-06) & (0.138)  &  (0.222)  & (0.139)  & (0.162)  & (0.164)   &  (0.033)\\
		\bottomrule
		\multicolumn{10}{p{430pt}}{\footnotesize{Notes: 1,565 observations. Significance levels: ${}^\dag$: 10\%, \, ${}^\ast$: 5\%, \, ${}^{\ast\ast}$: 1\%.}}
	\end{tabular}\par
	\medskip
	\label{tab:Earn}
\end{table}

\section{Proofs}
\label{app:proofs}

\subsection{Notation and preliminaries}

For any set $ \mathcal{G} $, let us denote by $ l^{\infty}(\mathcal{G}) $ the collection of all uniformly bounded real functions on $ \mathcal{G} $ equipped with the supremum norm $ \left\| f \right\|_{\mathcal{G}} = \sup_{x\in\mathcal{G}}\left| f(x) \right|  $.  Denote by $L^2(F)$  the square integrable space with respect to the measure associated with $F$, and let $ \left\| \cdot \right\|_{F,2} $ be the corresponding norm. We let $N(\epsilon,\mathcal{T},L_2(F)) $ denote the minimal number of $ \epsilon$-balls  with respect to $ \left\| \cdot \right\|_{F,2} $ needed to cover $ \mathcal{T} $.  An $ \epsilon$-bracket (with respect to $F$) is a pair of real functions $(l, u)$ such that $l\leq u$ and $ \left\|u-l\right\|_{F,2} \leq \epsilon$. Then, for any set  of real functions $ \mathcal{M}$, we let $N_{[]}(\epsilon,\mathcal{M},L_2(F))$ denote the minimum number of $ \epsilon$-brackets needed to cover $\mathcal{M}$. We denote by $\mathcal{G}= \left(\cup_{r\geq 1}\mathcal{G}_r\right)$. For $x \in \R^d$, $d>1$, we denote by $\left\|x \right\|_{\infty} = \max_{j=1,\dots,d} \abs{x}$.

\medskip
For a sequence of random variable $ (U_n)_{n\in\N} $ and a set $ \mathcal{F}_0 $, we say that $ U_n = O_P(1) $ uniformly in $ F \in \mathcal{F}_0 $ if for any $ \epsilon >0 $ there exist $ M >0 $ and $ n_0 >0 $ such that $ \sup_{F\in\mathcal{F}_0} \ \mathbb{P}_F\left( \abs{U_n} > M \right) < \epsilon $ for all $ n> n_0 $. Similarly we say that $ U_n = o_P(1) $ uniformly in $ F \in \mathcal{F}_0$ if for any $ \epsilon >0 $, $ \sup_{F\in\mathcal{F}_0} \ \mathbb{P}_F\left( \abs{U_n} > \epsilon \right) \to 0 $.

\medskip
Finally, we add stars to random variables whenever we consider their bootstrap versions, as with $T^*$ versus $T$. We define $o_{P^*} $ and $ O_{P^*}$ as above, but conditional on $ \left(\widetilde{Y}_i,D_i, X_i\right)_{i=1...n}$. Convergence in distribution conditional on $ \left(\widetilde{Y}_i,D_i, X_i\right)_{i=1...n}$ is denoted by $\to_{d^*}$.

\subsection{Proof of Lemma \ref{lem:reformulation}}

Under H$_0$, there exist  $Y', \psi'$ and $\mathcal{I}'$ such that $Y'\sim Y$, $\psi'\sim \psi$,  $\sigma(\psi')\subset \mathcal{I}'$ and $\E(Y'|\mathcal{I}')=\psi'$. Then, by the law of iterated expectations,
$$\E[Y'|\psi'] =\E\left[\E\left[Y'\middle|\mathcal{I}'\right]\middle|\psi'\right] = \E\left[\psi'|\psi'\right]=\psi'.$$
Conversely, if there exists $(Y',\psi')$ such that $Y'\sim Y$, $\psi'\sim \psi$ and $\E[Y'|\psi'] =\psi'$, let $\mathcal{I}'=\sigma\left(\psi'\right)$. Then
$\psi'= \E\left[Y'|\psi'\right]=\E\left[Y'|\mathcal{I}'\right]$ and H$_0$ holds.

\subsection{Proof of Theorem \ref{prop:equiv}}

(i) $\Leftrightarrow$ (iii). By Strassen's theorem \citep[][Theorem 8]{strassen1965}, the existence of $(Y,\psi)$ with margins equal to $F_Y$ and $F_\psi$ and such that $\E\left[Y|\psi\right]=\psi$ is equivalent to $\int f dF_\psi \leq \int fdF_Y$ for every convex function $f$. By, e.g., Proposition 2.3 in \cite{gozlan2015characterization}, this is, in turn, equivalent to (iii).

\medskip
(ii) $\Leftrightarrow$ (iii). By Fubini-Tonelli's theorem,
$\int_{-\infty}^y F_{Y}(t)dt = \E\left[\int_{-\infty}^y \indic\{t\geq Y\}dt\right] = \E\left[(y-Y)^+\right].$ The same holds for $\psi$. Hence, $\Delta(y)\geq 0$ for all $y \in \R $ is equivalent to $\E\left[\left(y-Y\right)^+\right] \geq  \E\left[\left(y-\psi\right)^+\right]$ for all $y\in \R$. The result follows.

\subsection{Proof of Proposition \ref{prop:1}}

First, by Jensen's inequality, we obtain
$$\E[(y_0-Y)^+|\psi] \geq (y_0-\E(Y|\psi))^+ = (y_0 - \psi)^+.$$
Moreover, $\Delta(y_0)=0$ implies that $\E((y_0-Y)^+) = \E((y_0-\psi)^+)$. Hence, almost surely, we have
$$\E[(y_0-Y)^+|\psi] = (y_0-\psi)^+.$$
Equality in the Jensen's inequality implies that the function is affine on the support of the random variable. Therefore, for almost all $u$, we either have $\Supp(Y|\psi=u) \subset [y_0, \infty)$ or $\Supp(Y|\psi=u) \subset (-\infty,y_0]$. Because $\E\left[Y|\psi\right]=\psi$, $\Supp(Y|\psi=u) \subset [y_0, \infty)$ for almost all $u>y_0$ and $\Supp(Y|\psi=u) \subset (-\infty,y_0]$ for almost all $u<y_0$. Then, for all $\tau \in (0,1)$, $F^{-1}_{Y|\psi}(\tau|u)\geq y_0$ for almost all $u\geq y_0$ and $F^{-1}_{Y|\psi}(\tau|u)\leq y_0$ for almost all $u\leq y_0$. Thus, for all $\tau \in (0,1)$, by continuity of $F^{-1}_{Y|\psi}(\tau|\cdot)$, $F^{-1}_{Y|\psi}(\tau|y_0)=y_0$. This implies that  $Y|\psi=y_0$ is degenerate.

\subsection{Proof of Proposition \ref{prop:equivX}}

We first prove that H$_{0X}$ is equivalent to the existence of $(Y',\psi')$ such that $DY'+(1-D)\psi' = \widetilde{Y}$, $D\indep(Y',\psi')|X$ and $\E((Y'|\psi',X)=\psi'$. First, under $H_{0X}$, there exists  $(Y',\psi',\mathcal{I}')$ such that $DY'+(1-D)\psi' = \widetilde{Y}$, $D\indep(Y',\psi')|X$, $\sigma(\psi',X)\subset \mathcal{I}'$ and $\E(Y'|\mathcal{I}')=\psi'$. Then
$$\E[Y'|\psi',X] =\E\left[\E\left[Y'\middle|\mathcal{I}'\right]\middle|\psi',X\right] = \E\left[\psi'|\psi',X\right]=\psi'.$$
Conversely, if there exists $(Y',\psi')$ such that $DY'+(1-D)\psi' = \widetilde{Y}$, $D\indep(Y',\psi')|X$ and $\E(Y'|\psi',X)=\psi'$, let  $\mathcal{I}'=\sigma\left(X',\psi'\right)$. Then
$\psi'= \E(Y'|\psi',X)=\E(Y'|\mathcal{I}')$ and H$_{0X}$ holds. The proposition then follows as Theorem \ref{prop:equiv}.

\subsection{Proof of Proposition \ref{prop:meas_err}} % (fold)
\label{sub:proof_of_proposition_ref_prop_meas_err}

For all $y$, $\xi \mapsto \E[(y- \psi - \xi)^+]$ is decreasing and convex. Then, because $F_{\xi_\psi}$ dominates at the second order $F_{\xi_Y+\eps}$, we have
$$\int \E\left[(y- \psi - \xi)^+\right] dF_{\eps+\xi_Y}(\xi) \geq \int \E\left[(y - \psi - \xi)^+\right] dF_{\xi_\psi}(\xi).$$
As a result, for all $y$, we obtain
\begin{align*}
\E\left[\left(y-\widehat{Y}\right)^+\right] = & \int \E\left[\left(y - \psi  - \eps -\xi_Y\right)^+| \eps+\xi_Y=\xi\right]dF_{\eps+\xi_Y}(\xi) \\
= & \int \E\left[(y -\psi - \xi)^+\right]dF_{\eps+\xi_Y}(\xi) \\
\geq & \int \E\left[(y - \psi - \xi)^+\right]dF_{\xi_\psi}(\xi) \\
= & \E\left[(y - \widehat{\psi})^+\right].
\end{align*}
Moreover, $\E\left(\widehat{Y}\right)=\E\left(\widehat{\psi}\right)$. By Theorem \ref{prop:equiv}, $\widehat{Y}$ and $\widehat{\psi}$ satisfy H$_0$.

% subsection proof_of_proposition_ref_prop_meas_err (end)

\subsection{Proof of Theorem \ref{prop:asympt_size}}

{\bf (i)} This is a particular case of Proposition \ref{prop:asympt_size_gen} below, with $q(Y,c_0) = Y$. The proof is therefore omitted.

\medskip
{\bf (ii)} We show that equality holds for $F_0\in \mathcal{F}_0$ satisfying the conditions stated in (ii). The proof is divided in three steps. We first prove convergence in distribution of $T$ to $S$ defined below, and conditional convergence of $T^*$ towards the same limit. Then we show that the cdf $H$ of $S$ is continuous and strictly increasing in the neighborhood of its quantile of order $1-\alpha$, for any $\alpha \in (0,1/2)$. The third step concludes.

\medskip
{\bf 1. Convergence in distribution of $T$ and $T^*$.}

\medskip
Let us introduce some notation. Let $K_{j,j}$ ($j\in \{1,2 \}$) be the $j$-th diagonal element of the covariance kernel $K$,  $ \mathcal{S}: \left(\nu,K\right) \mapsto (1-p) \left(-\nu_{1}/K_{1,1}^{1/2}\right)^{+2} + p \left(\nu_{2}/K_{2,2}^{1/2}\right)^2 $,  $q(r)=\left(r^2 + 100\right)^{-1} (2r)^{-d_X}$, and  \[ \nu_{n,F_0}(y,g) = \df{1}{\sqrt{n}}\sum_{i=1}^n  \mathrm{Diag}\left(\mathbb{V}_{F_0}\left(\widetilde{Y}\right)\right)^{-1/2} \left( m\left(D_i,\widetilde{Y}_{i}, X_i, g,y\right) - \E_{F_0}\left[m\left(D_i,\widetilde{Y}_{i}, X_i, g,y\right)\right]  \right). \]
Finally, we define $k_{n,F_0}(y,g) = \sqrt{n} \mathrm{Diag}\left(\mathbb{V}_{F_0}\left(\widetilde{Y}\right)\right)^{-1/2} \E_{F_0}\left[m\left(D_i,\widetilde{Y}_i,X_i,g,y \right)\right] $,
\begin{align*}
K_{n,F_0}(y,g,y',g') &=\mathrm{Diag}\left(\mathbb{V}_{F_0}\left(\widetilde{Y}\right)\right)^{-1/2}   \widehat{\text{Cov}}\left( \sqrt{n} \overline{m}_n(y,g), \sqrt{n} \overline{m}_n(y',g') \right)\mathrm{Diag}\left(\mathbb{V}_{F_0}\left(\widetilde{Y}\right)\right)^{-1/2}, \\
\overline{K}_{n,F_0}(y,g,y',g') &= K_{n,F_0}(y,g,y',g')  + \epsilon  \mathrm{Diag}\left(\mathbb{V}_{F_0}\left(\widetilde{Y}\right)\right)^{-1/2}\mathrm{Diag}\left( \widehat{\mathbb{V}}\left(\widetilde{Y} \right)       \right) \mathrm{Diag}\left(\mathbb{V}_{F_0}\left(\widetilde{Y}\right)\right)^{-1/2},
\end{align*}
and use the notations $K_{n,F_0}(y,g) = K_{n,F_0}(y,g,y,g) $ and $\overline{K}_{n,F_0}(y,g) = \overline{K}_{n,F_0}(y,g,y,g) $.

\medskip
We have, by definition of $T$,
\[ T = \sup_{y\in \mathcal{Y}} \sum_{(a,r): r\in\{1,...,r_n\}, a\in A_r} q(r)\mathcal{S}\left(\nu_{n,F_0}(y,g_{a,r}) + k_{n,F_0}(y,g_{a,r}), \overline{K}_{n,F_0}(y,g_{a,r}) \right). \]
To characterize the distribution of $T$ (resp. $T^*$), we first prove the convergence of $\nu_{n,F_0}$ and $K_{n,F_0}(y,g_{a,r})$ (resp. $ \nu_{n,F_0}^*$ and $K^*_{n,F_0}(y,g_{a,r})$). For those purposes, we use a class of functions which is a general form taken by $m_1$ defined in \eqref{eq:mn}, namely, for any $0<N_1<M_1$,
\begin{align*}
 \mathcal{M}_0 &= \{ f_{y,\phi_1,\phi_2,g}\left(\widetilde{y},x,d\right)=\left(d\phi_1 \left(y- \widetilde{y}\right)^+ - (1-d)\phi_2 \left(y- \widetilde{y}\right)^+   \right)g(x), \; \\ & \hspace{5cm} (y, \phi_1,\phi_2,g)\in \mathcal{Y}\times [N_1, M_1]^2 \times \mathcal{G}\}.
\end{align*}
Remark that $\mathcal{M}_0$ is a particular case of classes $\mathcal{M}$ defined in \eqref{eq:class} below. Then, by the proof of Proposition \ref{prop:asympt_size_gen} below, Assumptions PS1 and PS2 in AS are satisfied. Thus, the assumptions of Lemma D.2 in AS hold as well. This entails that Assumptions PS4 and PS5 in AS hold. Namely, there exists a Gaussian process $\nu_{F_0}$ such that
\begin{enumerate}
\item[-] $ \nu_{n,F_0} \to_{d} \nu_{F_0}$ and  $\nu_{n,F_0}^* \to_{d^*}  \nu_{F_0}$;
\item[-] For all $r \in \N$ and $ (y,g) \in \mathcal{Y}\times \mathcal{G}_r $, $\overline{K}_{n,F_0}(y,g) \to_P K_{F_0}(y,g)+\epsilon I_2$ and $K_{n,F_0}^*(y,g) \to_{P^*} K_{F_0}(y,g)+\epsilon I_2$, where $I_2$ is the $2\times 2$ identity matrix.
\end{enumerate}
Moreover, letting $ k_{F_0}(y,g) $ denote the limit in probability of $k_{n,F_0}(y,g)$, we have $k_{F_0}(y,g)= 0 $ if $ (y,g)\in\mathcal{L}_{F_0} $ and $ \infty $ otherwise. Note that by assumption, the set $ \mathcal{L}_{F_0} $ is nonempty.

\medskip
Thus, using (D.11) in the proof of Theorem D.3. in AS, which is based on the uniform continuity of the function $\mathcal{S}$ in the sense of Assumption S2 therein, we have, under $F_0$,
\begin{align*}
T & \to_{d}  \sup_{y\in \mathcal{Y}} \sum_{(a,r)\in A_r\times\N } \mathcal{S}\left(\nu_{F_0}(y,g_{a,r}) +  k_{F_0}(y,g_{a,r}), K_{F_0}(y,g_{a,r}) + \epsilon I_2 \right) \\
& = S := \sup_{y\in \mathcal{Y}} \sum_{(a,r): (y,g_{a,r}) \in\mathcal{L}_{F_0} } q(r) \mathcal{S}\left(\nu_{F_0}(y,g_{a,r}), K_{F_0}(y,g_{a,r})  + \epsilon I_2  \right),
\end{align*}
where the equality follows by definition of $\mathcal{S}$ and $k_{F_0}(y,g)$.
\medskip
Similarly, using Assumption PS5 and (D.11) in AS, replacing $T$ by $T^*$ and quantities $\nu_{n,F_0}(y,g_{a,r}) $ and  $K_{n,F_0}(y,g_{a,r}) $ by their bootstrap counterparts (see the proof of Lemma D.4 in AS) we have  $ T^*\to_{d^*} S$.

\medskip
{\bf 2. The cdf $H$ of $S$ is continuous and strictly increasing in the neighborhood of any of its quantile of order $1-\alpha>1/2$.}

\medskip
First, the cdf $H$ of $S$ is a convex functional of the Gaussian process $\nu_{F_0}$. Then, as in the proof of Lemma B3 in \cite{andrews2013inference}, we can use Theorem 11.1 of \cite{davydov1998local} p.75 to show that $H$ is continuous and strictly increasing at every point of its support except $\underline{r}=\inf\{ r \in \R: H(r) >0\}$. Moreover, for any $r>0$,
\begin{align*}
H(r)& \geq \mathbb{P}\left( \sup_{y\in \mathcal{Y}} \sum_{(a,r): (y,g_{a,r}) \in\mathcal{L}_{F_0} } q(r) \mathcal{S}\left(\nu_{F_0}(y,g_{a,r}), K_{F_0}(y,g_{a,r}) + \epsilon I_2  \right) < r  \right)\\
& \geq  \mathbb{P}\left( \sup_{ j\in\{1,2\}, (y,a,r): (y,g_{a,r}) \in  \mathcal{L}_{F_0}}  \left|(K_{2,F_0,j,j}(y,g_{a,r})  + \epsilon )^{-1/2} \nu_{F_0,j}(y,g_{a,r}) \right| < \frac{\sqrt{r/2}}{Q} \right) \\
& >0,
\end{align*}
where $Q = \sum_{(a,r): (y,g_{a,r}) \in\mathcal{L}_{F_0} } q(r) < \infty$ and we use Problem 11.3 of \cite{davydov1998local} p.79 for the last inequality. This yields $r>\underline{r}$ and $H$ is continuous and strictly increasing on $(0,\infty)$.

\medskip
Then, we show that for any $\alpha \in (0,1/2)$, the quantile of order $1-\alpha$  of the distribution of $ S $ is positive. By assumption, there exists $(y_0,g_0) \in\mathcal{L}_{F_0}$ such that either $K_{F_0,11}(y_0,g_0)>0$ or $K_{F_0,2}(y_0,g_0)>0$. This yields
\begin{align}
\mathbb{P}\left(S >0\right)& = 1 - \mathbb{P}\left( \sup_{y\in \mathcal{Y}} \sum_{(a,r): (y,g_{a,r}) \in\mathcal{L}_{F_0} } q(r) \mathcal{S}\left(\nu_{F_0}(y,g_{a,r}), K_{F_0}(y,g_{a,r} ) + \epsilon I_2  \right) =0 \right) \nonumber \\
& \geq 1 -  \mathbb{P}\left(\nu_{F_0,1}(y_0,g_0) \leq 0, \,
\nu_{F_0,2}(y,g_0) = 0   \right) \nonumber \\
& \geq 1- \min\left\{\mathbb{P}\left(\nu_{F_0,1}(y_0,g_0) \leq 0\right),
\mathbb{P}\left(\nu_{F_0,2}(y_0,g_0) = 0\right)\right\} \nonumber \\
& \geq  1/2. \label{eq:proba_Spos}
\end{align}
The first inequality holds by definition of the supremum and because $\mathcal{S}$ is nonnegative. To obtain the last inequality, note that either $\nu_{F_0,1}(y_0,g_0) $ is non-degenerate, in which case the first probability is $1/2$ (since $\nu_{F_0,1}(y_0,g_0)$ is normal with zero mean), or $\nu_{F_0,2}(y_0,g_0) $ is non-degenerate, in which case the second probability is $0$.

\medskip
Finally, using that $H$ is strictly increasing on $(0,\infty)$, \eqref{eq:proba_Spos} ensures that any quantile of $S$ of order $1-\alpha$ with $\alpha \in [0,1/2)$ is positive. Hence, $H$  is continuous and strictly  increasing in the neighborhood of any such quantiles.

\medskip
{\bf 3. Conclusion.}

\medskip
Using  $ T^*\to_{d^*} S$ in distribution, Step 2 and Lemma 21.2 in \cite{van2000asymptotic}, we have that for $\eta >0$, $ c^*_{n,\alpha} \to_{d^*}  c(1-\alpha + \eta ) + \eta $, where $c(1-\alpha +\eta)$ is the $(1-\alpha +\eta)$-th quantile of the distribution of $S$. Because $ T\to_{d} S$ and $H$ is continuous at $c(1-\alpha +\eta) + \eta>0$, we obtain that
$$ \underset{\eta \to 0}{\lim} \ \underset{n \to \infty}{\limsup} \  \mathbb{P}_{F_0}\left( T > c^*_{n,\alpha}  \right) = \alpha.$$
Combined with the inequality of Part (i) above, this yields the result.

\medskip
{\bf (iii)} This results follows from Theorem E.1 in AS. First, Assumption SIG2 in AS holds for $\sigma_{F}^2 = \mathbb{V}_{F}\left(\widetilde{Y}\right) $, following the proof of Lemma 7.2 (b) under Assumption \ref{hyp:regul}-(ii). Second, Assumptions PS4 and PS5 are satisfied using the point (ii) above. Third, Assumptions CI, MQ, S1, S3, S4 in AS are also satisfied by construction of the statistic $T$. Thus, Theorem E.1 in AS yields the result. \hfill $\square$

\subsection{Proof of Proposition \ref{prop:asympt_size_gen}} % (fold)
\label{sub:proof_of_proposition_ref_prop_asympt_size_gen}

 We introduce $\E_{F,c}=\E_{F}\left[m\left(D_i,\widetilde{Y}_{c,i}, X_i, g,y\right)\right]$ and
\begin{align*}
 \nu_{n,F}(y,g)& =\df{1}{\sqrt{n}}\sum_{i=1}^n  \mathrm{Diag}\left(\widehat{\mathbb{V}}_F\left(\widetilde{Y}_{\widehat{c}}\right)\right)^{-1/2} \left( m\left(D_i,\widetilde{Y}_{\widehat{c},i}, X_i, g,y\right) - E_{F,\widehat{c}} \right),\\
\overline{\nu}_{n,F}(y,g)& =\df{1}{\sqrt{n}}\sum_{i=1}^n \mathrm{Diag}\left(\mathbb{V}_F\left(\widetilde{Y}_{c_0}\right)\right)^{-1/2} \left( m\left(D_i,\widetilde{Y}_{c_0,i}, X_i, g,y\right) - E_{F,c_0}    \right).
\end{align*}
The proof is based on Theorem 5.1 in AS, hence we have to check that the corresponding assumptions  PS1, PS2, and SIG1 hold.  Namely, we have to ensure that
\begin{enumerate}
\item[-] \textbf{PS1}: for all sequence $F \in \mathcal{F}$ and all $(d,y',x,g,y,c)\in \{0,1\}\times\mathcal{Y}\times [0,1]^{d_X}\times\mathcal{G}_r\times \mathcal{Y}\times \mathcal{C}_s\left([0,1]^{d_X}\right) $
$$   \left| \dfrac{m(d,y',x,g,y)}{\mathbb{V}_{F}\left(\widetilde{Y}_{c,i}\right)} \right| \leq M(d,y',x,g,y) \ \text{and} \  \E_{F}\left[M\left(D_i,\widetilde{Y}_{c,i},X_i,g,y\right)^{2+\delta} \right] \leq C < \infty,  $$
 where $\delta >0$ and for some function $ M$;
\item[-] \textbf{PS2}:  for all sequence $F_n \in \mathcal{F}$, the i.i.d triangular array of processes
\begin{align*}
\mathcal{T}^0_{n} & = \bigg\{  \dfrac{m\left(D_i,\widetilde{Y}_{n,c(X_{n,i})},X_{n,i},g,y\right)}{\mathbb{V}_{F_n}\left(\widetilde{Y}_{n,c(X_{n,i})}\right)} ,\ (c,y,g)\in \mathcal{C}_s\left([0,1]^{d_X}\right)\times \mathcal{Y}\times \mathcal{G}, \  i\leq n, \ n\geq 1\bigg\}	
\end{align*}
is manageable with respect to some envelope function $ U_1$ \citep[see][p.38 for the definition of a manageable class]{pollard1990empirical};
\item[-]\textbf{SIG1}: for all $\zeta >0$, $\sup_{F \in \mathcal{F}, c \in \mathcal{C}_s\left([0,1]^{d_X} \right)} \mathbb{P}\left( \left| \widehat{\mathbb{V}}_F\left(\widetilde{Y}_{i,c}\right)/\mathbb{V}_F\left(\widetilde{Y}_{i,c}\right) -1 \right| > \zeta  \right) \to 0$.
\end{enumerate}

We proceed in two steps, to handle the fact that $c_0$ and  $\mathrm{Diag}\left(\mathbb{V}_F\left(\widetilde{Y}_{c_0}\right)\right)^{-1/2}$ are estimated:
\begin{enumerate}
\item We first show that	
\begin{align}
\sup_{F\in\mathcal{F}_0}\underset{g\in \cup_{r\geq 1}\mathcal{G}_r, y\in\mathcal{Y}}{\sup} \left\| \nu_{n,F}(y,g)- \overline{\nu}_{n,F}(y,g)  \right\|_{\infty} = & o_{P}(1), \label{eq:approx1}\\
\sup_{F\in\mathcal{F}_0}\underset{g\in \cup_{r\geq 1}\mathcal{G}_r, y\in\mathcal{Y}}{\sup}\left\| \nu^*_{n,F}(y,g)- \overline{\nu}^*_{n,F}(y,g)  \right\|_{\infty} = & o_{P^*}(1).\label{eq:approx2}
\end{align}
\item Next, we show that $ m $ satisfies assumptions PS1, PS2, and that SIG1 in AS also holds for $\sigma_{F}^2 = \mathbb{V}_{F}\left(\widetilde{Y}_{c_0}\right) $, where $F \in\mathcal{F}$ and $\widehat{\sigma}_{n}^2 = n^{-1}\sum_{i=1}^n\left(\widetilde{Y}_{\widehat{c}, i} - n^{-1} \sum_{j=1}^n\widetilde{Y}_{\widehat{c}, j} \right)^2 $. \end{enumerate}

{\bf 1. Proof of \eqref{eq:approx1}-\eqref{eq:approx2}}

\medskip
We apply the uniform version over $ F\in\mathcal{F}_0 $ of Theorem 3 in \cite{chen2003estimation} to a general class of functions to which pertain the moment condition $m$ (see \eqref{eq:mn}, with $\widetilde{Y}$ replaced here by $\widetilde{Y}_{c}= D q\left(\widetilde{Y}, c \right)+  (1-D)\psi$ and without the moment equality $m_2$).  Hence, it suffices to verify that Assumptions (3.2) and (3.3) of Theorem 3 in \cite{chen2003estimation} are satisfied. Let us introduce,  for any $0 <N_1 < M_1$,  the classes of functions
\begin{align}\label{eq:class}
\mathcal{M}_1=& \left\{f_{c,y,\phi,g}\left(\widetilde{y},x\right)=\phi \left(y- q\left(\widetilde{y},c(x)\right)\right)^+g(x),(c,y, \phi,g)\in \mathcal{C}_s\left([0,1]^{d_X}\right)\times \mathcal{Y}\times [N_1, M_1]\times \mathcal{G} \right\}, \\
\mathcal{M}_2=& \left\{f_{c,y,\phi,g}\left(\widetilde{y},x\right)=\phi \left(y- \widetilde{y}\right)^+g(x), \; (c,y, \phi,g)\in \mathcal{C}_s\left([0,1]^{d_X}\right)\times \mathcal{Y}\times [N_1, M_1]\times \mathcal{G} \right\}, \notag \\
\mathcal{M}=& \{  f_{c,y,\phi_1,\phi_2,g}\left(\widetilde{y},x,d\right) = (dg_{c,y,\phi_1,g} -(1-d) q_{c,y,\phi_2,g})\left(\widetilde{y},x\right),\ g \in \mathcal{M}_1, \  q\in \mathcal{M}_2, \notag \\
& \hspace{6cm} (c,y, \phi_1,\phi_2,g)\in \mathcal{C}_s\left([0,1]^{d_X}\right)\times \mathcal{Y}\times [N_1, M_1]^2\times \mathcal{G}  \}. \notag
\end{align}
Note that $\phi_1$, $\phi_2$, and $c$ in the class $\mathcal{M}$ denote components of $m$ that are estimated.

\medskip
Consider the space $\mathcal{C}_s\left([0,1]^{d_X}\right)\times \mathcal{Y}\times [N_1,M_1]^2\times \mathcal{G}$ equipped with the norm  $$\left\| (c,y,\phi_1,\phi_2,g)\right\| = \max\left\{\left\| c \right\|_{[0,1]^{d_X}}, \abs{y},\abs{\phi_1},\abs{\phi_2}, \left\| g \right\|_{[0,1]^{d_X}}\right\}.$$
For $ v =(c,y,\phi_1,\phi_2,g), v'=(c',y',\phi_1',\phi_2',g') \in \mathcal{C}_s\left([0,1]^{d_X}\right)\times \mathcal{Y}\times [N_1,M_1]^2 \times \mathcal{G}$ and $(\widetilde{y},x,d) \in \mathcal{Y}\times [0,1]^{d_X}\times\{0,1\}$, we have, by the triangular inequality and Assumptions \ref{hyp:gen}-(i) and \ref{hyp:gen}-(v),
\begin{align*}
\abs{ f_{v}\left(\widetilde{y},x,d\right) - f_{v'}\left(\widetilde{y},x,d\right) } \leq & \abs{ g_{c,y,\phi_1,g}\left(\widetilde{y},x\right) - g_{c',y',\phi'_1,g'}\left(\widetilde{y},x\right) } \\
& + \abs{ q_{c,y,\phi_2,g}\left(\widetilde{y},x\right) - q_{c',y',\phi'_2,g'}\left(\widetilde{y},x\right) }\\
 \leq &  (M+M_0)\left( \abs{ \phi_1 - \phi_2' } + \abs{ \phi_2 - \phi_2' } \right) \\
&+ 2M_1\left[\left| y - y' \right| + \left| q\left(\widetilde{y},c(x)\right) - q\left(\widetilde{y},c'(x)\right)    \right|\right] \\
 & + 2 M_0M_1 \big[\left| \indic\left\{ q(\widetilde{y},c(x))\leq y    \right\} - \indic\left\{q(\widetilde{y},c(x))\leq y'  \right\}  \right|\\
& \hspace{1.7cm} + \left|\indic\left\{q\left(\widetilde{y},c(x)\right)\leq y'\right\} - \indic\left\{q\left(\widetilde{y},c'(x)\right)\leq y'\right\}\right| \\ & \hspace{1.7cm} + \left| g(x) - g'(x)\right|  \big].
\end{align*}
Denote by $K_q>0$ the Lipschitz constant of $q(\widetilde{y},.)$. Then, by convexity of $x\mapsto x^2$, we obtain
\begin{align*}
\dfrac{1}{7}\abs{ f_{v}\left(\widetilde{y},x,d\right) - f_{v'}\left(\widetilde{y},x,d\right) }^2 \leq &  (M+M_0)^2\left( \abs{ \phi_1 - \phi_1' }^2 + \abs{ \phi_2 - \phi_2' }^2 \right)\\
& + 4M_1^2\left[\left| y - y' \right|^2 + K_q \left\| c -c'\right\|_{[0,1]^{d_X}}^2\right] \\ & +  4(M_0M_1)^2 \big[\left| \indic\left\{ q(\widetilde{y},c(x))\leq y    \right\} - \indic\left\{q(\widetilde{y},c(x))\leq y'  \right\}  \right|  \\
&  \hspace{2cm}  +  \left|\indic\left\{q\left(\widetilde{y},c(x)\right)\leq y'\right\} - \indic\left\{q\left(\widetilde{y},c'(x)\right)\leq y'\right\}\right|\\
 & \hspace{2cm} + \left\| g - g'\right\|^2_{[0,1]^{d_X}}  \big].
\end{align*}
Fix $ \delta >0 $. If $\left\| v-v'\right\| \leq \delta$, this yields
\begin{align*}
\dfrac{1}{7}\abs{ f_{v}\left(\widetilde{y},x,d\right) - f_{v'}\left(\widetilde{y},x,d\right) }^2\leq
&  \delta^2\left(2(M+M_0)^2 + 4M_1^2(1+K_q) + 4(M_0M_1)^2\right) \\
& +  4(M_0M_1)^2 \big[ \indic\left\{ q(\widetilde{y},c(x))\leq y +\delta    \right\} - \indic\left\{q(\widetilde{y},c(x))\leq y-\delta  \right\}   \\
  &   \hspace{2cm} +  \left|\indic\left\{\widetilde{y}\leq q^I\left(y',c(x)\right)\right\} - \indic\left\{\widetilde{y}\leq q^I\left(y',c'(x)\right)\right\}    \right|\big].
\end{align*}
Next, by Assumption \ref{hyp:gen}-(iv), we obtain
\begin{align*}
 \E\left[ \indic\left\{ q\left(\widetilde{Y},c(X)\right)\leq y +\delta    \right\} - \indic\left\{q\left(\widetilde{Y},c(X)\right)\leq y -\delta  \right\} \right]& =  F_{q\left(\widetilde{Y},c(X)\right)}\left( y +\delta\right) - F_{q\left(\widetilde{Y},c(X)\right)}\left( y -\delta\right) \\
&\leq 2 \overline{Q}_{2}\delta.
\end{align*}
Finally, we have
\begin{align*}
&\E\left[ \left|\indic\left\{Y\leq q^I\left(y',c(X)\right)\right\} - \indic\left\{\widetilde{y}\leq q^I\left(y',c'(X)\right)\right\}\right| \right]\\
\leq &  \E\left[ \indic\left\{Y\leq q^I\left(y',c(X)\right) -Q_{F,2} \delta\right\} - \indic\left\{\widetilde{y}\leq q^I\left(y',c(X)\right) +  Q_{F,2} \delta\right\}\right]  \\
\leq & \E\left[  F_{Y|X}\left(q^I\left(y',c(X)\right) - Q_{q^I} \delta \middle| X\right)  -F_{Y|X}\left(q^I\left(y',c(X)\right) +Q_{q^I} \delta \middle| X\right)     \right]\\
\leq & 2Q_{F,1} Q_{q^I}\delta,
\end{align*}
where $ Q_{q^I} $ is the Lipschitz constant of $ q^I $. Thus, by Assumption \ref{hyp:gen}, there exists $ Q >0 $  such that
\begin{equation}\label{eq:Q}
\sup_{F\in\mathcal{F}_0} \E\left[ \underset{\left\|v-v'\right\|\leq \delta }{\sup}  \left|  f_{v}\left(\widetilde{Y},X,D\right) - f_{v'}\left(\widetilde{Y},X,D\right) \right|^2 \right] \leq Q\delta.
\end{equation}
Therefore the class $ \mathcal{M} $ satisfies Condition (3.2) of Theorem 3 in \cite{chen2003estimation} uniformly in $ F\in\mathcal{F}_0 $. Moreover, the class $ \mathcal{G}$ is manageable and thus Donsker \citep[see Lemma 3 in][]{andrews2013inference}. Finally, by Remark 3 (ii) in \cite{chen2003estimation},  $\mathcal{C}_s\left([0,1]^{d_X}\right)$ is also Donsker. Then, $ \mathcal{C}_s\left([0,1]^{d_X}\right) $, $ \mathcal{Y} $, $ [N_1,M_1] $, and $ \mathcal{G} $ satisfy Condition (3.3) of Theorem 3 in \cite{chen2003estimation}. The result follows by Theorem 3 in \cite{chen2003estimation}.

\medskip
{\bf 2. $m$ satisfies PS1 and PS2 of AS and SIG1 of AS also holds for $\sigma_{F}^2$ and $\widehat{\sigma}_{n}^2$.}

\medskip
From Assumption \ref{hyp:gen} (iii) and the proof of Lemma 7.2 (a) in AS,  PS1 is satisfied replacing $ B $ by $\max(M,M_0)$ in the proof of Lemma 7.2-(a) in AS.

\medskip
We now show that PS2 in AS also holds. As the result is uniform over $\mathcal{F}_0$, we have to consider sequences for the cdfs $F_n$ of $( D_{n,i},Y_{n,i},X_{n,i})_{i=1...n}$ (with $F_n\in \mathcal{F}_0$). We also define
\begin{align*}
	\widetilde{Y}_{n,c(X_{n,i})} & = D_{n,i} q\left(Y_{n,i},c(X_{n,i})\right) + (1-D_{n,i})\psi_{n,i}, \\
	W_{n,i} & = \frac{D_{n,i}}{\E_{F_n}\left[D_{n,i}\right]}- \frac{1 - D_{n,i}}{\E_{F_n}\left[1- D_{n,i}\right]}, \\
	\sigma_{F_n}^2 & = \mathbb{V}_{F_n}\left(\widetilde{Y}_{n,c(X_{n,i})}\right).
\end{align*}
Note that by Assumption \ref{hyp:regul} (iii), $\sigma_{F_n}^2 \geq \overline{\sigma} >0$ for all $F_n\in \mathcal{F}$. Let $(\Omega,\mathbb{F}, F_n)$ be a probability space and let $\omega$ denote a generic element in $\Omega$. Showing Assumption PS2 in AS then boils down to prove  that for any $0< N_1  < M_1:= 1/\inf_F \sigma_F^2$, the i.i.d triangular array of processes
\begin{align*}
\mathcal{T}_{1,n,\omega} & = \bigg\{ W_{n,i} \phi \left(y-\widetilde{Y}_{n,c(X_{n,i})}\right)^+g(X_{n,i}),\ (c,y,\phi,g)\in \mathcal{C}_s\left([0,1]^{d_X}\right)\times \mathcal{Y} \times [N_1,M_1]\times \mathcal{G},  \\
& \qquad   i\leq n,n \geq 1    \bigg\}
\end{align*}
is manageable with respect to some envelope function $ U_1$. Lemma 3 in \cite{andrews2013inference} shows that the processes
$\left\{  g(X_{n,i}), \  g \in\mathcal{G}, \  i\leq n,n \geq 1   \right\} $ are manageable with respect to the constant function 1. Then, using Lemma D.5 in AS, it remains to show that
\[ \mathcal{T}'_{1,n,\omega} = \left\{ W_{n,i} \phi \left(y-\widetilde{Y}_{n,c(X_{n,i})}\right)^+,\ (c,y,\phi)\in \mathcal{C}_s\left([0,1]^{d_X}\right)\times \mathcal{Y} \times [N_1,M_1],   i\leq n,n \geq 1    \right\}, \]
is manageable with respect to some envelope. For such an envelope, we can consider
$U_1'(\omega) = (M_0 + M)/(\overline{\sigma}\epsilon_0) $. We now prove the manageability of $\mathcal{T}'_{1,n,\omega}$. Let us define
\begin{align*}
\mathcal{M}' = & \left\{f_{c,y,\phi_1,\phi_2}\left(\widetilde{y},x,d\right)= d\phi_1 \left(y- q\left(\widetilde{y},c(x)\right)\right)^+ - (1-d)\phi_2 \left(y- \widetilde{y}\right)^+,  \right. \\
& \left. \; (c,y, \phi_1,\phi_2)\in \mathcal{C}_s\left([0,1]^{d_X}\right)\times \mathcal{Y}\times [N_1,M_1]^2  \right\}.
\end{align*}
Reasoning as for the class $\mathcal{M}$ defined in \eqref{eq:class}, and using the last equation of the proof of Theorem 3 in \cite{chen2003estimation}, p.1607, we have that for $ \epsilon>0$, 
$$ N_{[\cdot]}\left( \epsilon,\mathcal{M}', \norm{\cdot}_2  \right) \leq N\left(\epsilon', [N_1,M_1]^2, \abs{\cdot} \right)\times N\left( \epsilon', \mathcal{Y}, \abs{\cdot} \right) \times N\left( \epsilon', \mathcal{C}_s\left([0,1]^{d_X}\right), \norm{\cdot}_{[0,1]^{d_X}} \right),$$
with $\epsilon' =(\epsilon/(2Q))^2$ and $Q$ defined in \eqref{eq:Q}. Using Theorem 2.7.1 page 155 in \cite{van1996weak}, there exists a constant $Q_2$ depending only on $s$, $d_X$, and $[0,1]^{d_X}$ such that
\begin{equation*}\label{eq:control_eq1}
\ln\left( N \left(\epsilon', \mathcal{C}_s([0,1]^{d_X}), \norm{\cdot}_{[0,1]^{d_X}} \right)\right) \leq  Q_2 \epsilon'{}^{-d_X/s}.
\end{equation*}
Moreover, because $ \mathcal{Y} $ and $ [N_1,M_1] $ are compact subsets of two Euclidean spaces, there exist $Q_3$, $Q_4$  such that
\begin{equation}
N\left( \epsilon', [N_1,M_1] ^2, \abs{\cdot} \right)\leq  Q_3 \epsilon'^{-4} \ \text{and} \  N\left( \epsilon', \mathcal{Y}, \abs{\cdot} \right) \leq Q_4 \epsilon'^{-2} . \label{eq:control_eq2}
\end{equation}
This yields
\begin{equation}
 \ln\left( N_{[\cdot]}\left( \epsilon,\mathcal{M}', \norm{\cdot}_2  \right)\right)\leq (  6+Q_2) \max\left( -\ln(\epsilon'),\epsilon'^{-  d_X/s }\right)+  \ln(Q_3 Q_4)	.
	\label{eq:ineq_entr_crochet}
\end{equation}
 Let $ \odot $ denote element-by-element product and $ \mathcal{D}\left(\epsilon\left|\alpha \odot U_1'(\omega) \right|,\alpha \odot \mathcal{T}'_{1,n,\omega} \right) $ denote random packing numbers. By (A.1) in \citeauthor{andrews1994empirical} (1994, p.2284), we have
\begin{align}
\underset{\omega \in \Omega,n\geq 1, \ \alpha \in \R_+^n}{\sup} \mathcal{D}\left(\epsilon\left|\alpha \odot U_1'(\omega)\right|,\alpha \odot \mathcal{T}'_{1,n,\omega} \right)  & \leq \underset{F \in \mathcal{F}_0}{\sup}  N\left(\df{ \epsilon}{2},\mathcal{M}',  \norm{\cdot}_2\right) \notag\\
& \leq \underset{F \in \mathcal{F}_0}{\sup}  N_{[\cdot]}\left( \epsilon,\mathcal{M}',  \norm{\cdot}_2\right),\label{def:pack}
\end{align}
where the second inequality follows as in e.g., \citeauthor{van1996weak} (1996, p.84). Then, \eqref{eq:ineq_entr_crochet} ensures (see Definition 7.9 in \cite{pollard1990empirical}, p.38) that $$\underset{\omega \in \Omega,n\geq 1, \ \alpha \in \R_+^n}{\sup} \mathcal{D}\left(\epsilon\left|\alpha \odot U_1'(\omega) \right|,\alpha \odot \mathcal{T}'_{1,n,\omega} \right) \leq \lambda(\epsilon),$$ where
$ \lambda(\epsilon)= \exp\left( (6+Q_2) \max\left(- 2\ln\left(\epsilon/(2Q)\right),\left(\epsilon/(2Q)\right)^{-2 d_X/s }\right) +  \ln(Q_3 Q_4) \right)$. Moreover, by using $\sqrt{a+b} \leq \sqrt{a} + \sqrt{b}$
for all $a,b \geq0$,
\begin{align*}
\int_{0}^1\sqrt{\ln(\lambda(\epsilon))} d\epsilon &\leq  \sqrt{6+Q_2} \int_{0}^1 \left[\max\left(- 2\ln\left(\epsilon/(2Q)\right),\left(\epsilon/(2Q)\right)^{-2 d_X/s }\right)\right]^{1/2}  d\epsilon  +\sqrt{ \ln(Q_3 Q_4)}\\
& < \infty.
\end{align*}
Thus, $\mathcal{T}'_{1,n,\omega}$ hence $\mathcal{T}_{1,n,\omega}$ are manageable. Therefore, $m$ satisfies PS2 in AS.

\medskip
Finally, in order to show that SIG1 in AS is satisfied, we use Assumption \ref{hyp:gen} (iii) and follow the proof of Lemma 7.2 (b) in AS where we replace $ Y $ by $ q(Y,c(X)) $ and $ B $ by $\max(M,M_0)$. The result follows.

% subsection proof_of_proposition_ref_prop_asympt_size_gen (end)

\subsection{Proof of Proposition \ref{prop:imposs_result}} % (fold)

Hereafter, we let $[\underline{\psi},\overline{\psi}]$ (resp. $[\underline{y},\overline{y}]$) denote the support of $\psi$ (resp. of $Y$). As in Lemma \ref{lem:reformulation}, $H_{0SK}$ holds if and only if there exists a pair of random variables $(Y',\psi')$ and $c$ such that  $Y' \sim Y$, $\psi' \sim \psi$ and $\E\left[Y'\middle|\psi'\right]=Q_c(\psi')$. Now, if $Q_c$ is strictly increasing on $[\underline{\psi},\overline{\psi}]$, we have $\E\left[Y'\middle|\psi'\right]=Q_c(\psi')$ if and only if $\E\left[Y'\middle|Q_c(\psi')\right]=Q_c(\psi')$. In view of Theorem \ref{prop:equiv}, the latter is equivalent to $F_Y$ being a mean-preserving spread of $F_{Q_c(\psi')}$. Therefore, the proposition holds if for any $\eta>0$, there exists $K$, $c\in \R^{K+1}$ and $F$ such that (i) $Q_c$ is strictly increasing on $[\underline{\psi},\overline{\psi}]$; (ii)  $\sup_{y\in\R} |F_\psi(y)-F(y)|<\eta$; (iii) $F_Y$ is mean-preserving spread of $F_{Q_c(\widetilde{\psi})}$, with $\widetilde{\psi}\sim F$.

\medskip
Fix $\eta>0$. Since $F_Y$ is continuous on $[\underline{y},\overline{y}]$, it is uniformly continuous on this set. Hence, there exists $\eta'$ such that 
\begin{equation}
|y-y'|<\eta'  \Rightarrow |F_Y(y)-F_Y(y')|<\eta.	
	\label{eq:unif_cont}
\end{equation}
By assumption, $F_Y^{-1}\circ F_\psi$ is increasing and continuous. Then, by  Theorem 9 in \cite{mulansky1998interpolation}, there exists a sequence $(P_n)_{n\in\N}$ of increasing polynomials on $[\underline{\psi},\overline{\psi}]$ satisfying $P_n(\underline{\psi})=\underline{y}$ and $P_n(\overline{\psi})=\overline{y}$ and converging uniformly to $F_Y^{-1}\circ F_\psi$. Hence, there exists  $P_{n_0}$ such that 
\begin{equation}
\sup_{y\in [\underline{\psi},\overline{\psi}]} |P_{n_0}(y) - F_Y^{-1}\circ F_\psi(y)|<\eta'.
	\label{eq:approx_pol}
\end{equation}
Let $K$ be the degree of $P_{n_0}$ and $c\in \R^K$ denote the vector of coefficients of $P_{n_0}$, so that $Q_c=P_{n_0}$. $Q_c$ is a non-constant polynomial, which is increasing on $[\underline{\psi},\overline{\psi}]$. Hence, its derivative vanishes a finite number of times and $Q_c$ is actually strictly increasing. Hence, Condition (i) above holds. Moreover, combining \eqref{eq:approx_pol} with \eqref{eq:unif_cont}, we obtain
$$\sup_{y\in [\underline{\psi},\overline{\psi}]} |F_Y \circ Q_c(y) - F_\psi(y)|<\eta.$$ 
Now, let $F:=F_Y \circ Q_c$ on $[\underline{\psi},\overline{\psi}]$, $F(y):=0$ for all $y< \underline{\psi}$ and  $F(y):=1$ for all $y> \overline{\psi}$. Then $F$ is continuous and increasing, with limit 0 and 1 respectively at $-\infty$ and $\infty$. Thus, it is a cdf and Condition (ii) above holds. Finally, let $\widetilde{\psi}\sim F$. We have, for any $y\in [\underline{y},\overline{y}]$,
$$P\left(Q_c(\widetilde{\psi})\leq y\right)= F \circ Q_c^{-1}(y) = F_Y(y).$$
This implies that $F_{Q_c(\widetilde{\psi})}$ is a mean-preserving spread of $F_Y$. The result follows.

% subsection proof_of_proposition_ref_prop_imposs_result (end)

\subsection{Proof of Proposition \ref{prop:reg_meas_err}} % (fold)
\label{sub:proof_of_proposition_ref_prop_reg_meas_err}

1. We consider for that purpose $(\psi^*, \xi^*_\psi,\xi^*_Y,\eps^*)\sim \mathcal{N}(m,\Sigma)$, potentially different from the true  $(\psi, \xi_\psi,\xi_Y,\eps)$, and let
\begin{align*}
\psh^* & =\psi^*+\xi^*_\psi,\\	
\Yh^* & = \psi^* + \eps^* + \xi^*_Y.
\end{align*}
We then fix $(m,\Sigma)$  so that the DGP satisfies all the  restrictions specified in the propositions, and in particular, $(\V(\Yh^*), \V(\psh^*),\Cov(\Yh^*,\psh^*))=(\V(\Yh), \V(\psh),\Cov(\Yh,\psh))$. First, letting $m=(m_1,m_2,m_3,m_4)'$, we impose $m_2=m_3=m_4=0$, and set all the non-diagonal terms of $\Sigma$, except $\Sigma_{23}=\Cov(\xi^*_\psi,\xi^*_Y)$, equal to zero. Then $(\Yh^*,\psh^*,\psi^*)$ satisfy \eqref{eq:meas_errors} and RE hold (considering $\mathcal{I}=\sigma(\psi^*)$ and $Y^*=\psi^* + \eps^* $). We fix below $\Sigma_{22}\in [0,\V(\psh)]$. Then let  $\Sigma_{11}=\V(\psh)-\Sigma_{22}$ and $\Sigma_{33}=\V(\Yh)-\V(\psh)+\Sigma_{22}$ and $\Sigma_{44}=0$, so that $(\V(\Yh^*), \V(\psh^*))=(\V(\Yh), \V(\psh))$. Also, because $\V(\Yh)>\V(\psh)$, $\V(\xi^*_\psi) <\V(\xi^*_Y+\eps^*)$ and $F_{\xi^*_\psi}$ dominates at the second order $F_{\xi^*_Y+\eps^*}$.

\medskip
Now, we fix $\Sigma_{22}$. Let $a=\V(\Yh) - \V(\psh)$ and $c=\Cov(\Yh - \psh, \psh)$. Then, by Cauchy-Schwarz inequality,
$$c^2 \leq  \V(\psh) \V(\Yh -\psh) = \V(\psh) (a-2c).$$
This means that there exists $\sigma^2 \in [0,\V(\psh)]$ such that
\begin{equation}
c^2 \leq \sigma^2 (a-2c).	
	\label{eq:ineg_for_cov}
\end{equation}
Let $\Sigma_{22}=\sigma^2$ and $\Sigma_{23}=c+ \Sigma_{22}$. Then, by construction,
\begin{align*}
	\Cov(\Yh^*,\psh^*) &=  \Sigma_{11}+ \Sigma_{23} \\
	& =  \V(\psh)-\Sigma_{22} + \Sigma_{22} + c \\
	& = \Cov(\Yh , \psh).
\end{align*}
Moreover, in view of \eqref{eq:ineg_for_cov} and by definition of $\Sigma_{22}$ and $\Sigma_{33}$,
\begin{align*}
\Sigma_{23}^2 & = c^2 +2 c \Sigma_{22} + \Sigma_{22}^2 \\
& \leq (a-2c)\Sigma_{22} + 2 c \Sigma_{22} + \Sigma_{22}^2 \\
&= \Sigma_{33}\Sigma_{22}.
\end{align*}
In other words, $\Sigma$ is a proper covariance matrix.

\medskip
2. Let $\lambda=\V(\psi)/\sigma^2_{\xi_\psi}$. If \eqref{eq:meas_errors} and RE hold, $\Cov(\xi_\psi, \eps+\xi_Y)\geq 0$ and $\lambda\geq \underline{\lambda}$, we obtain
\begin{align*}
\beta - 1 = & \frac{\Cov(\Yh-\psh,\psh)}{\V(\psh)} \\
= & \frac{\Cov(\eps+\xi_Y-\xi_\psi,\xi_\psi)}{\sigma^2_{\xi_\psi}(1+\lambda)} \\
\geq & - \frac{1}{1+\underline{\lambda}}.
\end{align*}
The result follows.

% subsection proof_of_proposition_ref_prop_reg_meas_err (end)

\subsection{Proof of Proposition \ref{prop:equiv_bounds}} % (fold)
\label{sub:proof_of_proposition_ref_prop_equiv_bounds}

We first prove that if $\E[\psi_L]\leq \E[Y]\leq \E[\psi_U]$, there exists a unique $F^*\in\mathcal{F}_B$ such that $\delta_{F^*}=0$. First, suppose that $F^b \neq F^{b'}$ and, without loss of generality, $b > b'$. Then $\psi^{b}\leq \psi^{b'}$, implying that $F^b(y)\leq F^{b'}(y)$ for all $y$. Moreover, the inequality is strict for at least one $y$. As a result, $\E(\psi^{b})>\E(\psi^{b'})$. In other words, there is at most one $F^*\in\mathcal{F}_B$ such that $\delta_{F^*}=0$. If $\E[\psi_L]= \E[Y]$ or $\E[\psi_U]= \E[Y]$, such a solution also exists by taking $b=-\infty$ and $b=\infty$, respectively. Now, suppose that $\E[\psi_L]< \E[Y]< \E[\psi_U]$. For all $\infty>b> b'>-\infty$,
\begin{align*}
\psi^b - \psi^{b'} = & \left(\psi_U - \max(\psi_L,b')\right)\indic\{\psi_U\in [b',b)\}+(b-b')\indic\{\psi_L < b', \psi_U\geq b\} \\
& + (b-\psi_L) \indic\{\psi_L \in [b',b), \psi_U\geq b\}.	
\end{align*}
As a result, $|\psi^b - \psi^{b'}|\leq |b-b'|$. This implies that $\widetilde{\delta}:b\mapsto \E[\psi^b]$ is continuous. Moreover, $\lim_{b\rightarrow-\infty} \widetilde{\delta}(b)=\E[\psi_L]<\E(Y)$ and $\lim_{b\rightarrow\infty} \widetilde{\delta}(b)=\E[\psi_U]>\E(Y)$. By the intermediate value theorem, there exists $b^*$ such that $\widetilde{\delta}(b^*)= \E(Y)$. Hence, there exists $F^*\in\mathcal{F}_B$ such that $\delta_{F^*}=0$. The first part of Proposition \ref{prop:equiv_bounds} follows.

\medskip
Let us turn to the second part of the proposition. First, if (ii) holds, there exists $b_0\in\overline{\R}$ such that $F^*=F^{b_0}$. Then, by construction and Theorem \ref{prop:equiv}, $Y$ and $\psi^{b_0}$ satisfy H$_0$. Moreover, $F^{b_0} \in [F_{\psi_U},F_{\psi_L}]$. Therefore, H$_{0B}$ holds as well.

\medskip
Now, let us prove that (i) implies (ii). Let us denote by $ \mathcal{D}$ the set of all the cdfs for $\psi$ such that $ \text{H}_{0B} $ holds. By Theorem \ref{prop:equiv}, these are cdfs $F$ satisfying $ F_{\psi_U} \leq F \leq F_{\psi_L}$, $\delta_F=0$ and dominating at the second order $F_{Y}$. We show below that all $ F\in \mathcal{D} $ are dominated at the second order by $F^* $. Then, because $F_{\psi_U} \leq F^* \leq F_{\psi_L}$ and $\int ydF^*(y)=\int ydF_Y(y)$, $\mathcal{D}$ is not empty only if $F^*$ dominates at the second order $F_Y$. The result then follows by Theorem \ref{prop:equiv}.

\medskip
Thus, we have to show that for all $ t\in\R $,
\begin{equation}
F^* = \argmin_{F_{\psi} \in\mathcal{D}} \int_{-\infty}^{t} F_{\psi}(y) dy.	\label{eq:F_psi_c0_min}
\end{equation}
First, if $F^*=F^{-\infty}$, we have for all $F\neq F^*$, $F(y)\leq F_{\psi_L}(y)=F^*(y)$ for all $y$, with strict inequality for some $y$. Then $\delta_F>\delta_{F^*}=0$ and $\mathcal{D}=\{F^*\}$, implying that \eqref{eq:F_psi_c0_min} holds. Similarly, \eqref{eq:F_psi_c0_min} holds if $F^*=F^{\infty}$.

\medskip
Suppose now that $F^*=F^{b_0}$ for some $b_0\in \R$. Because $  F_{\psi_U}(y)\leq F_\psi(y)$ for all $y <b_0 $ and all $ F_{\psi} \in \mathcal{D} $, \eqref{eq:F_psi_c0_min} holds for all $t<b_0$. We now prove that \eqref{eq:F_psi_c0_min} holds also for $t\geq b_0$. First suppose that $t \geq \max(b_0,0)$. For all $ F_{\psi} \in \mathcal{D} $, $\int y dF_{Y}(y)=\int y dF_{\psi}(y)dy$. As a result, by Fubini's theorem,
\begin{align*}
&-\int_{-\infty}^{0}F^*(y) dy +  \int_{0}^{t}\left(1 - F^*(y)\right) dy + \int_{t}^{\infty}\left(1  - F^*(y)\right) dy
\\ &=-\int_{-\infty}^{0}F_{\psi}(y) dy +\int_{0}^{t} \left(1 - F_{\psi}(y) \right)dy + \int_{t}^{\infty} \left(1 - F_{\psi}(y)\right) dy.
\end{align*}
Because $  F_{\psi}\leq F_{\psi_L}=F^*$ on $[b_0,\infty]$,
this implies that
$$-\int_{-\infty}^{0}F^*(y) dy +  \int_{0}^{t}\left(1 - F^*(y)\right) dy \geq -\int_{-\infty}^{0}F_{\psi}(y) dy +\int_{0}^{t} \left(1 - F_{\psi}(y) \right)dy $$
and thus \eqref{eq:F_psi_c0_min} holds for $t \geq \max(b_0,0)$. Now, if $b_0 <0$ and $t \in (b_0, 0)$, we have
\begin{align*}
&-\left(\int_{-\infty}^{t}F^*(y) dy +  \int_{t}^{0} F^*(y)dy \right)+ \int_{0}^{\infty}\left(1  - F^*(y)\right) dy
\\ &=-\left(\int_{-\infty}^{t}F_{\psi}(y) dy +\int_{t}^{0}  F_{\psi}(y) dy \right)+ \int_{0}^{\infty} \left(1 - F_{\psi}(y)\right) dy.
\end{align*}
Using again $  F_{\psi}\leq F_{\psi_L}=F^*$ on $[t,\infty)$ yields
\[  - \int_{t}^{0} F^*(y)dy + \int_{0}^{\infty}\left(1  - F^*(y)\right) dy  \leq - \int_{t}^{0}  F_{\psi}(y) dy + \int_{0}^{\infty} \left(1 - F_{\psi}(y)\right) dy. \]
Therefore, the result also follows in this case.

% subsection proof_of_proposition_ref_prop_equiv_bounds (end)

\end{document}